\documentclass[tradiabstract]{aa} 
\usepackage{graphicx}
\usepackage{txfonts}
\usepackage{natbib}
\usepackage{longtable}
\usepackage{lscape}
\usepackage{multirow}

\bibpunct{(}{)}{;}{a}{}{,}

\newcommand{\mum}{\,$\mu$m}

\newcommand{\Msun}{$M_{\odot}$}

\newcommand{\SII}{[S\,II]}

\newcommand{\CaII}{Ca\,II}
\newcommand{\HeI}{He\,I}

\newcommand{\nodata}{...}
\newcommand{\accunit}{$M_{\odot}$~yr$^{-1}$}
\newcommand{\rev}{}
\newcommand{\newrev}{}
\newcommand{\newnewrev}{}

\newcommand{\fig}{Fig.}
\newcommand{\Mjup}{$M_{\rm{Jup}}$}
\newcommand{\Mdotacc}{$\dot M_{\rm{acc}}$}

\begin{document}

\title{Young stars in $\epsilon$ Cha and their disks\thanks{Based on observations performed at ESO's La Silla-Paranal observatory under programme 076.C-0470}: disk evolution in sparse associations}

\author{M.~Fang\inst{1,2} \and  R.~van~Boekel\inst{1}  \and J.~Bouwman\inst{1}  \and Th.~Henning\inst{1} \and W.~A.~Lawson\inst{3} \and A.~Sicilia-Aguilar\inst{4}}
\institute{Max-Planck Institute for Astronomy, K\"onigstuhl 17,
           D-69117 Heidelberg, Germany
           \and
           Purple Mountain Observatory and Key Laboratory for Radio Astronomy, 2 West Beijing Road, 210008 Nanjing, China
           \and
           School of Physical, Environmental \& Mathematical Sciences, University of New South Wales, Canberra ACT 2600, Australia       \and
          Departamento de F\'{\i}sica Te\'{o}rica, Universidad Aut\'{o}noma de Madrid, Cantoblanco 28049, Madrid, Spain
    }

   \date{Received 27 November 2011; accepted 25 September 2012}

 \abstract{The nearby young stellar association $\epsilon$ Cha has an estimated age of 3--5\,Myr, making it an ideal laboratory to study the disk dissipation process and  provide empirical constraints on the timescale of planet formation.}
{We wish to complement existing optical and near-infrared data of the $\epsilon$ Cha association, which provide the stellar properties of its members, with mid-infrared data that probe the presence, geometry, and mineralogical composition of protoplanetary disks around individual stars.}
{We combine the available literature data with our Spitzer IRS spectroscopy and VLT/VISIR imaging data. We use  proper motions to refine the membership of $\epsilon$\,Cha. Masses and ages of individual stars are estimated by fitting model atmospheres to the optical and near-infrared photometry, followed by placement in the HR-diagram. The Spitzer IRS spectra are analyzed using the two-layer temperature distribution  spectral decomposition method.}
{Two stars previously identified as members, CXOU\,J120152.8 and 2MASS\,J12074597, have proper motions that are very different from those of the other stars. But other observations suggest that the two stars are still young and thus might still be related to $\epsilon$ Cha. HD\,104237C is the lowest mass member of $\epsilon$ Cha with an estimated mass of $\sim$13--15 Jupiter masses. The very low mass stars USNO-B120144.7 and 2MASS\,J12005517 show globally depleted spectral energy distributions, pointing at strong dust settling. 2MASS\,J12014343 may have a disk with a very specific inclination, where the central star is effectively screened by the cold outer parts of a flared disk, but the 10\mum \ radiation of the warm inner disk can still reach us. We find that the disks in sparse stellar associations are dissipated more slowly than those in denser (cluster) environments. We detect C$_{2}$H$_{2}$  rovibrational band around 13.7\mum\  on the IRS spectrum of USNO-B120144.7. We find strong signatures of grain growth and crystallization in all $\epsilon$\,Cha members with 10\mum \ features detected in their IRS spectra. We combine the dust properties derived in the $\epsilon$\,Cha sample with those found  using identical or similar methods in the MBM\,12, Coronet, $\eta$\,Cha associations, and  in the cores-to-disks legacy program. We find that disks around low-mass young stars show a negative radial gradient in the mass-averaged grain size and mass fraction of crystalline silicates. A positive correlation exists between the mass-averaged grain sizes of amorphous silicates and the accretion rates if the latter is above $\sim$10$^{-9}$\,\accunit, possibly indicating that those disks are sufficiently turbulent to prevent grains of several microns in size to sink into the disk interior.}
{}

 \maketitle

\section{Introduction}

The observational characterization of the structure and evolution of circumstellar disks is key to our understanding of the disk dissipation and planet formation processes. Nearby pre-main sequence (PMS) associations at $\sim$50-300\,pc are well suited for detailed investigations of young stars and their disks, since their members  can be observed with high signal-to-noise ratio (S/N) throughout the electromagnetic spectrum. Several PMS associations have been studied with the Spitzer Space Telescope, e.g., MBM\,12, $\epsilon$\,Cha,  and $\eta$\,Cha \citep[][]{2009A&A...497..379M,2005ApJ...634L.113M,2006ApJ...653L..57B,2009ApJ...701.1188S} using the infrared array camera \citep[IRAC,][]{2004ApJS..154...10F}, the multiband imaging photometer for Spitzer \citep[MIPS,][]{2004ApJS..154...25R}, and the infrared spectrograph \citep[IRS,][]{2004ApJS..154...18H}. The Spitzer data allow characterization of the disks around low-mass stars up to radii of $\sim$10\,AU. The observed disk frequencies in the MBM\,12 and $\eta$\,Cha associations suggest that disk dissipation proceeds more slowly in these sparse PMS associations compared to denser environments \citep[see, e.g.,][]{2012A&A...539A.119F}. MBM\,12 has a disk frequency of $\approx${\newnewrev 67\%} at an age of 2\,Myr \citep{2009A&A...497..379M}, while in the 5-10\,Myr old $\eta$\,Cha association, 40-50\% of the stars still retain a disk \citep{1999ApJ...516L..77M,2005ApJ...634L.113M,2009ApJ...701.1188S}. In the latter cluster, binarity was shown to play an important role in protoplanetary disk evolution \citep{2006ApJ...653L..57B}.

Silicates are an important dust component in protoplanetary disks. The most convincing evidence that silicates are present in the protoplanetary disks is the strong ``10\mum\ feature'' in mid-infrared spectra of Herbig Ae/Be and T~Tauri stars \citep{1985ApJ...294..345C,2000ApJ...534..838N,2001A&A...375..950B,2003A&A...401..577B,2004Natur.432..479V,2005A&A...437..189V}. This feature is emitted by silicate grains with sizes of up to several microns that reside in the optically thin surface layers of the disk \citep[][]{1997A&A...318..879M,1997ApJ...490..368C}. The shape and the strength of the 10\mum\ silicate feature were found to be correlated and this was interpreted as evidence for grain growth in disks \citep{2005A&A...437..189V,2006ApJ...639..275K,2008ApJ...683..479B}. The Spitzer IRS  has provided spectra of many young objects, covering the wavelength range from 5.3 to 38\mum. Analysis of the spectral features in the Spitzer IRS spectra provides constraints on the chemical composition and grain sizes \citep[see, e.g.,][for a recent review]{2010ARA&A..48...21H}. The dust properties derived from Spitzer IRS spectra of the members of the MBM\,12 and $\eta$\,Cha associations suggest that dust processing in the disks occurs very early and that radial mixing is not efficient \citep{2009A&A...497..379M,2009ApJ...701.1188S}.

Because the $\eta$\,Cha association is much older than MBM\,12, it is highly desirable to study an association at an intermediate age in order to draw a more complete picture of disk evolution in sparse stellar associations. With an estimated age of 3--5\,Myr \citep{2003ApJ...599.1207F}, the $\epsilon$\,Cha association is well suited for this purpose. Located at a distance of $\sim$114\,pc, the $\epsilon$\,Cha association was discovered by \citet{2003ApJ...599.1207F}, who identified nine members. The most massive member, $\epsilon$\,Cha\,AB, has a spectral type of B9. \citet{2004ApJ...616.1033L} found three new members of $\epsilon$\,Cha from a survey covering a region with a radius of 0\fdg5 around $\epsilon$\,Cha\,AB.  At 3--5\,Myr, the $\epsilon$\,Cha\ association is younger than the $\eta$\,Cha group \citep[5-10\,Myr,][]{1999ApJ...516L..77M,2004ApJ...609..917L}. \citet{2009MNRAS.400L..29L} provide a model-independent way to rank the ages of the nearest PMS associations by employing gravity-sensitive spectral features in optical spectra. They confirmed that the $\epsilon$\,Cha association is younger than the $\eta$\,Cha and TW~Hya associations.

In this paper, we will investigate the disks surrounding the $\epsilon$\,Cha members in terms of overall (geometry) evolution and dust mineralogy. We arrange this paper as follows. In Section~\ref{Sec:data}, we describe the observations and data reduction. In Section~\ref{Sec:result} we present our results, which are then discussed. We summarize our efforts in Section~\ref{Sec:summary}.

\section{Observations, data reduction, and analysis}\label{Sec:data}
\subsection{Targets}\label{Sec:targets}

There are 12 stars reported as the members of $\epsilon$\,Cha in the literature \citep{2003ApJ...599.1207F,2004ApJ...616.1033L}. We collected the photometric and spectroscopic data of these stars that are available in the literature and list these in Tables~\ref{Tab:optical_photometry}, \ref{Tab:infrared_photometry}, and \ref{Tab:acc}. For the ten stars whose spectral types have been estimated, we performed spectral energy distribution (SED) fits in the optical and near-infrared spectral range to derive the luminosity of each star and its line-of-sight extinction, using the method described in \citet{2009A&A...504..461F}. We take a model atmosphere spectrum with a fixed effective temperature corresponding to the observed spectral type and generate synthetic photometry with two free parameters: the visual extinction $A_{V}$ and the stellar angular diameter $\theta$. Both fit parameters are fine tuned to minimize the residuals between synthetic and observed optical and near-infrared photometry. We adopt a standard extinction law \citep{1989ApJ...345..245C} with a total-to-selective extinction ratio of $R_{V}$\,=\,3.1. We used all the available optical photometry and the near-infrared $J$- and $H$-band photometry. The bolometric luminosity of each star is then obtained by integrating the de-reddened model spectrum and adopting a distance of 114\,pc. For the $\epsilon$\,Cha\,AB, which is a double system, we estimate the luminosity of the A component from its $V$-band magnitude \citep[5.34\,mag,][]{2008hsf2.book..757T}, which was corrected for binarity. The resulting extinctions and luminosities are listed in Table~\ref{Tab:acc}.

\begin{table*}
\caption{Optical and near-infrared photometry for members in the $\epsilon$\,Cha association. Column 5: the $B$-band photometry of member ID\#2,5 comes from the Tycho-2 Catalogue \citep{2000A&A...355L..27H}; the $B$-band photometry for member ID\#10,11,12 is obtained with their $V$-band photometry and  $B-V$ colors \citep{2008MNRAS.389.1461L}. Column 6: the $V$-band photometry of member ID\#1, 6, 7, 8, 9 comes from \citet{2003ApJ...599.1207F}; the $V$-band photometry of member ID\#2,5 comes from the Tycho-2 Catalogue \citep{2000A&A...355L..27H}; the $V$-band photometry of member ID\#10,11,12 is obtained from its $I$-band photometry and $V-I$ colors \citep{2008MNRAS.389.1461L}. Column 7: the $R$-band photometry of member ID\#1,6,7,8,9 comes from \citet{2003ApJ...599.1207F}, whereas the $R$-band photometry of member ID\#10,11,12 is obtained with its $I$-band photometry and $R-I$ colors from \citep{2008MNRAS.389.1461L}. Column 8: the $I$-band photometry for member ID\#1,6,7,8,9 comes from \citet{2003ApJ...599.1207F}, whereas the $I$-band photometry for member ID\#10,11,12 is from the DENIS  survey \citep{1997Msngr..87...27E}. Columns 9, 10, 11: the photometry for member ID\#3,4,6,7 is from \citet{2004ApJ...608..809G}. The photometry for other members comes from the 2MASS survey \citep{2006AJ....131.1163S}.}\label{Tab:optical_photometry}             
\scriptsize
\centering                                      
\begin{tabular}{llcccccccccccccccccccccccc}          
\hline\hline                        
(1)   & (2)                 &(3)           &(4)             &(5)         &(6)           &(7)        &(8)       &(9)           &(10)        &(11)          \\
      &                     &RA            & DEC            &$B$           &$V$             &$R$         &$I$           &$J$            &$H$           &$K_{s}$       \\
ID    & Name                &(J2000)       & (J2000)        &(mag)       &(mag)         &(mag)     &(mag)       &(mag)        &(mag)       &(mag)         \\
\hline                                                                           
1&  CXOU J115908.2-781232   &  11:59:07.98 & -78:12:32.2    &\nodata     &16.99        &15.57       &13.83      &12.01        &11.45       &11.17         \\ 
2&  $\epsilon$\,Cha\,AB     &  11:59:37.53 & -78:13:18.9    &4.86         &4.91        &\nodata    &\nodata     &5.02         &5.04        &4.98          \\ 
3&  HD 104237C              &  12:00:03.89 & -78:11:31.0    &\nodata     &\nodata      &\nodata    &\nodata     &15.28        &14.85       &14.48      \\ 
4&  HD 104237B              &  12:00:04.76 & -78:11:34.8    &\nodata     &\nodata      &\nodata    &\nodata     &11.43        &10.27       &9.52        \\ 
5&  HD 104237A              &  12:00:05.21 & -78:11:34.4    &6.86        &6.62          &\nodata    &\nodata     &5.81         &5.25        &4.59          \\ 
6&  HD 104237D              &  12:00:08.39 & -78:11:39.2    &\nodata     &14.28        &13.09      &11.62       &10.53         &9.73        &9.67        \\ 
7&  HD 104237E              &  12:00:09.43 & -78:11:42.2    &\nodata     &12.08        &11.25      &10.28       &9.10          &8.05        &7.70        \\ 
8&  USNO-B120144.7-781926   &  12:01:44.42 & -78:19:26.8    &\nodata     &17.18        &15.61      &13.72       &11.68        &11.12       &10.78         \\ 
9&  CXOU J120152.8-781840   &  12:01:52.52 & -78:18:41.4    &\nodata     &16.78        &15.29      &13.52       &11.63        &11.04       &10.77         \\ 
10& 2MASS J12005517-7820296 &  12:00:55.17 & -78:20:29.7    &19.61       &17.85        &16.08      &14.00       &11.96        &11.40       &11.01         \\ 
11& 2MASS J12014343-7835472 &  12:01:43.43 & -78:35:47.2    &20.17       &18.55        &17.15      &15.96       &14.36        &13.38       &12.81         \\ 
12& 2MASS J12074597-7816064&  12:07:45.98 & -78:16:06.5     &17.68        &16.08        &14.74      &13.11       &11.55        &10.98       &10.67         \\ 
\hline                                                   
\end{tabular}
\end{table*}
\normalsize

\renewcommand{\tabcolsep}{0.17cm}
\begin{table*}
\caption{\label{Tab:infrared_photometry} Infrared photometry for members in the $\epsilon$\,Cha association. Columns 4,5,6: the photometry comes from \citet{2004ApJ...608..809G}. Column 7: the photometry comes from \citet{2010ApJS..186..111L}. Columns 8,9,10,11,12,13: the photometry is from the AKARI IRC Point Source Catalogue and FIS Bright Source Catalogue \citep{2010A&A...514A...1I,2010yCat.2298....0Y}. Columns 14,15,16,17: the photometry is from WISE Preliminary Release Source Catalog \citep{2010AJ....140.1868W}.}
\scriptsize
\centering                                      
\begin{tabular}{lllllllllllllllllllllllllllllllll}      
\hline\hline                        
(1)      & (2)          &(3)             &(4)    &(5)     &(6)      &(7)    &(8)    &(9)     &(10) &(11)  &(12)    &(13) &(14) &(15) &(16) &(17)\\
         &RA            & DEC            &$L'$     &$M$       &11.9\mum &[24]   & S9W   & L18W   &N60  &WIDE-S &WIDE-L &N160 &W1 &W2 &W3 &W4 \\ 
ID       &(J2000)       & (J2000)        &(mJy)  &(mJy)   &(mJy)    &(mJy)  &(mJy)  & (mJy)  &(mJy)&(mJy) &(mJy)   &(mJy) & (mag) &(mag) &(mag) &(mag)  \\
\hline                                                                                       
1        &  11:59:07.98 & -78:12:32.2    &\nodata&\nodata &\nodata  &\nodata&\nodata &\nodata &\nodata&\nodata&\nodata&\nodata &11.00 &10.77&10.58 &$>$8.73\\       
2        &  11:59:37.53 & -78:13:18.9    &\nodata&\nodata &\nodata  & 73.68 &561.4   &\nodata &\nodata&\nodata&\nodata&\nodata &5.81 &5.12 &5.10 &4.95\\       
5        &  12:00:05.12 & -78:11:34.7    &9795.  &9091.   &14167.   &\nodata&13479.7 &13000.1 &10036.2&8600.1 &5116.1 &4531.8 &4.01 &3.03&0.81 &-0.71 &\\       
6        &  12:00:08.30 & -78:11:39.6    & 45.   &\nodata &\nodata  &\nodata& \nodata& \nodata&\nodata&\nodata&\nodata&\nodata&\nodata&\nodata&\nodata&\nodata\\        
7        &  12:00:09.32 & -78:11:42.5    &481.   &329.    &498.     & 1055.2& \nodata& \nodata&\nodata&\nodata&\nodata&\nodata&\nodata&\nodata&\nodata&\nodata\\        
8        &  12:01:44.42 & -78:19:26.8    &\nodata&\nodata&\nodata   & 15.39 & \nodata& \nodata&\nodata&\nodata&\nodata&\nodata &10.36&9.83&8.34&6.74\\       
9        &  12:01:52.52 & -78:18:41.4    &\nodata&\nodata&\nodata   & 0.80  & \nodata& \nodata&\nodata&\nodata&\nodata&\nodata&10.60&10.35&10.04&$>$8.67\\       
10       &  12:00:55.17 & -78:20:29.7    &\nodata&\nodata&\nodata   & 15.54 & \nodata& \nodata&\nodata&\nodata&\nodata&\nodata&10.64&10.17&8.48&6.62\\       
11       &  12:01:43.43 & -78:35:47.2    &\nodata&\nodata&\nodata   & 55.38 & \nodata& \nodata&\nodata&\nodata&\nodata&\nodata&12.37&11.59&7.55&5.26\\       
12       &  12:07:45.98 & -78:16:06.5    &\nodata&\nodata&\nodata   & 0.88  & \nodata& \nodata&\nodata&\nodata&\nodata&\nodata&10.51&10.34&10.12&$>$8.72\\       
\hline   
\end{tabular}
\end{table*}
\normalsize                                          
 \renewcommand{\tabcolsep}{0.09cm}
\begin{table*}
\caption{Stellar parameters for members in the $\epsilon$\,Cha association.  Columns 4,5: the values come from \citet{2004ApJ...616.1033L}. 
{\newnewrev Columns 8,9: the masses and ages of the sources ID\#1,4,6,7,8,9,10,11,12 are estimated from \citet{2008ApJS..178...89D} and \citet{1998A&A...337..403B}, respectively. For the sources ID\#2A,5 with masses larger than 1.4\,\Msun, their masses and ages are estimated from  \citet{2008ApJS..178...89D}, and for the source ID\#3 with a mass less than 0.1\,\Msun, its mass and age are from  \citet{1998A&A...337..403B}.}  Column 10: The equivalent widths (EW) of H$\alpha$ line for ID\#1,2,5,6,7,8,9 come from \citet{2003ApJ...599.1207F}. The  EWs of H$\alpha$ line for ID\#10,11,12 are estimated from the optial spectra presented in \citet{2008MNRAS.389.1461L}. Negative values mean  H$\alpha$ line in emission. Column 11: W for weak T-Tauri star (WTTS), C for classical T-Tauri star (CTTS), and H for Herbig Ae/Be star.}\label{Tab:acc}             
\scriptsize
\centering                                      
\begin{tabular}{lllcccccccccccccc}          
\hline\hline                        
(1)    & (2)          &(3)          &(4)    &(5)    &(6)            &(7)    &(8)  &(9)      &(10)      &(11) &(12)    &(13)   &(14) &(15) &(16)\\
       &RA            & DEC         &       &$T_{\rm eff}$   &$L_{\rm bol}$    &$A_{V}$    &mass &age       &H$\alpha$&     &$\dot{M}$$_{acc}$& & $\mu_{\alpha}cos(\delta)$& $\mu_{\delta}$&\\
ID     &(J2000)       & (J2000)     & Spt   &(K)    &($L\odot$)      &(mag) &(\Msun) &(Myr)  &($\AA$)   &Class&(\accunit)         &Disk &(mas/yr)&(mas/yr)&Member\\    
\hline                                   
  1    &  11:59:07.98 & -78:12:32.2 & M4.75 & 3161  &0.027 & 0      &0.17/0.14    &9.6/7.1     &-6.2       &W  &\nodata    &N &-35.7$\pm$13.6&-5.8$\pm$13.6&Y\\
  2A   &  11:59:37.53 & -78:13:18.9 & B9    &10500  &99.7  & 0      &2.87/\nodata &2.7/\nodata  &+13        &\nodata &\nodata    &N&-35.1$\pm$1.5&4.0$\pm$1.9&Y\\
  3    &  12:00:03.60 & -78:11:31.0 &\nodata&\nodata&\nodata&\nodata&\nodata/0.015&\nodata/5&\nodata   &\nodata&\nodata&\nodata &\nodata&\nodata  &Y\\ 
  4    &  12:00:04.00 & -78:11:37.0 &\nodata&\nodata&\nodata&4.5&0.42/0.17&4&\nodata    &\nodata&\nodata&Y &\nodata&\nodata &Y\\
  5    &  12:00:05.12 & -78:11:34.7 &A7.75  &7648   &42.8   & 0.91   &2.53/\nodata &3.2/\nodata  &-20        &H  &4.17E-07  &Y &-36.9$\pm$1.4&-5.6$\pm$1.8&Y\\
  6    &  12:00:08.30 & -78:11:39.6 &M3.5   &3342   &0.12  & 0       &0.29/0.30 &3.0/3.8  &-3.9       &W  &\nodata   &N  &\nodata&\nodata&Y\\
  7    &  12:00:09.32 & -78:11:42.5 &K5.5   &4278   &0.92  & 1.0    &0.90/1.33  &2.1/5.9  &-4.5       &C  &8.59E-10  &Y &\nodata&\nodata&Y\\
  8    &  12:01:44.42 & -78:19:26.8 &M5     &3125   &0.036 & 0      &0.16/0.14  &5.1/4.5  &-23        &C  &1.49E-11  &Y&-31.5$\pm$7.7&0.5$\pm$7.7&Y\\
  9    &  12:01:52.52 & -78:18:41.4 &M4.75  &3161   &0.039 & 0      &0.18/0.15  &5.9/5.0  &-7.8       &W  &\nodata   &N&-28$\pm$7.7&-34$\pm$7.7&?\\
  10   &  12:00:55.17 & -78:20:29.7 &M5.75  &3024   &0.029 & 0      &0.11/0.10  &3.1/3.5  &-23.7      &C  &1.05E-11  &Y&-33.6$\pm$7.7&-2.2$\pm$7.7&Y\\
  11   &  12:01:43.43 & -78:35:47.2 &M2.25  &3524   &0.0038& 0      &0.27/0.49  &4    &-192.4     &C  &\nodata &Y&-43.1$\pm$8.8&-6.5$\pm$8.8&Y\\
  12   &  12:07:45.98 & -78:16:06.5 &M3.75  &3306   &0.042 & 0      &0.27/0.25  &12.5/11.4  &-7.9       &W  &\nodata &N&-68.7$\pm$7.8&-17.9$\pm$7.7&?\\
\hline                                             
 \end{tabular}
\end{table*}
\normalsize

\subsection{VISIR imaging}

{\rev In the  $\epsilon$\,Cha association, four sources (HD104237B-E) are located too close to the bright star HD104237A to be properly resolved with the Spitzer telescope. To characterize the disk properties of these sources, we observed them with the VLT imager and spectrometer for mid-infrared (VISIR) on the night of 2006 March 17 through the SiC filter, which transmits radiation between $\sim$10.7 and $\sim$12.9\mum.} We employed standard chopping and nodding techniques to eliminate  high instrumental and atmospheric background emission, with a chop throw of 8\arcsec in the north-south direction and a nod throw of 8\arcsec in the east-west direction. We assessed the quality of the individual chop half-cycles (here referred to as ``frames'') by fitting a 2D Gaussian function to the A component, which is detected at high S/N in each frame. We find that the quality of most frames is good, meaning the image of the A component is sharp and round. The median full width at half maximum of the 75\% best frames is 0\farcs33, and we shift-and-add these into our final image with sub-pixel accuracy, using shifts determined by cross correlation.

\begin{figure}[ht]
\centering
\includegraphics[width=0.7\columnwidth,angle=90]{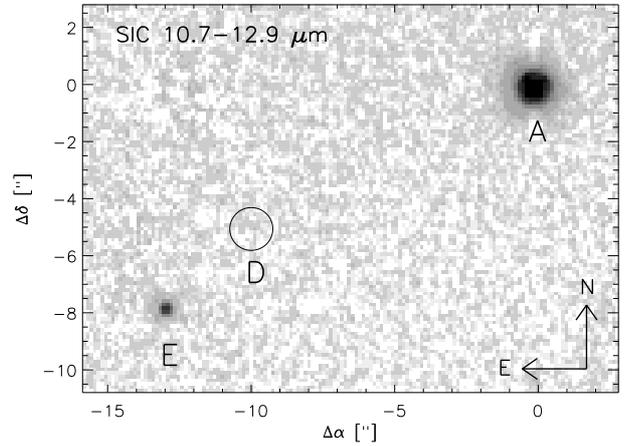}
\caption{VLT/VISIR image of the HD\,104237 region taken through the SIC filter, on a square root scaling from -0.1\% to +6.5\% of the peak flux of the HD\,104237A component. The position of the  HD\,104237D component is indicated with a circle, but there is no significant detection of this source. Coordinates are offsets in arcseconds with respect to HD\,104237A.}\label{Fig:VISIR}
\end{figure}

In \fig~\ref{Fig:VISIR}, we show our final VISIR image of the HD\,104237 system. The A and E components, corresponding to sources 3 and 6 in \citet{2004ApJ...608..809G}, respectively, are clearly detected. No sources are seen at the positions of the C and D components \citep[source numbers 1 and 5 in][]{2004ApJ...608..809G}. We did not attempt an absolute flux calibration of our VISIR data, but instead performed point-spread function photometry on the E component, using the A component as the reference star. We find an A/E flux ratio of 23.4$\pm$0.4. To set the absolute flux scale, we integrated the Spitzer spectrum of the A component over the transmission curve of the SIC filter. Assuming that the spectrum of the E component is iso-photonic, one obtains a flux of 551$\pm$10\,mJy at 12\mum\ for this source. At the position of the D component, we do not detect a source with confidence and derive a 3$\sigma$ upper limit of 18\,mJy at 12\mum, again assuming an iso-photonic spectrum.

\subsection{Spitzer IRS spectroscopy}
\label{Sec:IRS}

The $\epsilon$\,Cha members were observed with the Spitzer Space Telescope as part of a large program aimed at studying the evolution of circumstellar disks in nearby associations (GO proposal 20691, PI: Bouwman). We obtained 7--35\mum\, low-resolution (R=60--120) spectra with the IRS. The extracted spectra are based on the droopres products processed through the S18.7.0 version of the Spitzer data pipeline. Partially based on the SMART software package \citep{2004PASP..116..975H}, our data were further processed using spectral extraction tools developed by the Formation and Evolution of Planetary Systems (FEPS) Spitzer science legacy team \citep[see also ][]{2008ApJ...683..479B}. The spectra were extracted using a 6.0 pixel and 5.0 pixel fixed-width aperture in the spatial dimension for observations with the first order of the short- (7.5--14\mum) and the long-wavelength (14--35\mum) modules, respectively. The background was subtracted using associated pairs of imaged spectra from the two nod positions along the slit, also eliminating stray light contamination and anomalous dark currents. Pixels tagged by the data pipeline as being ``bad'' were replaced with a value interpolated from an 8-pixel perimeter surrounding the errant pixel. Low-level fringing at wavelengths $>$20\mum\ was removed using the {\sc irsfinge} package \citep{2003cdsf.conf..335L}. To remove any effect of pointing offsets, we matched orders based on the point-spread function of the IRS instrument, thereby correcting for possible flux losses. The spectra are calibrated using a spectral response function derived from multiple IRS spectra of the calibration star $\eta^{1}$~Doradus and a MARCS stellar model provided by the Spitzer Science Center. The spectra of the calibration target were extracted in the same way as our science targets. The relative errors between spectral channels within one order are dominated by the noise in each channel and not by the calibration. We estimate the relative flux calibration within a spectral order to be good to $\approx$1\% and the absolute calibration between different orders to be accurate to $\approx$3\%, the uncertainty being dominated by uncertainties in the scaling of the MARCS model.

We use the two-layer temperature distribution (TLTD) spectral decomposition routines developed by \citet{2009ApJ...695.1024J,2010ApJ...721..431J} to derive the mineralogical composition of the dust from the observed IRS spectra. {\rev The TLTD method is applicable to disks at most inclinations but does not work well for objects with a near edge-on inclination \citep{2009ApJ...695.1024J}. Hence, we can apply it to all sources in this survey, with the exception of 2MASS~J12014343 (see Sect.~\ref{Sec:edg_on}).} The spectrum emitted by the optically thin surface layers of the disk is approximated by the  expression
\begin{equation}
F_{\rm \nu}=F_{\rm \nu,cont}+\sum_{i=1}^{N}\sum_{j=1}^{M}D_{i,j}\kappa_{\rm i,j} \int_{T_{\rm a,max}}^{T_{\rm a,min}}\frac{2\pi}{d^{\rm 2}}B_{\rm \nu}(T)T^\frac{2-qa}{qa}dT,  
\end{equation}
where the different dust species in the model are denoted with subscripts $i$; for each species, we include different grain sizes, indicated with $j$ subscripts. $\kappa_{\rm i,j}$ denotes the mass absorption coefficient of dust species $i$ with grain size $j$. The abundances of each dust component are indicated by $D_{\rm i,j}$, while $B_{\rm \nu}(T)$ denotes the Planck function and $qa$ represents the exponent of the adopted power-law temperature distribution. The disk atmosphere has a range of temperatures between the integration boundaries $T_{\rm a}$, and $d$ denotes the distance between the star and Earth. $F_{\rm \nu,cont}$ denotes the continuum flux from the optically thick disk interior and is approximated by
\begin{eqnarray}
F_{\rm \nu,cont}=D_{\rm 0}\frac{\pi R^{2}_{\star}}{d^{2}}B_{\rm \nu}(T_{\rm \star})+D1\int_{T_{\rm r,max}}^{T_{\rm r,min}}\frac{2\pi}{d^{\rm 2}}B_{\rm \nu}(T)T^\frac{2-qr}{qr}dT \nonumber \\
+D2\int_{T_{\rm m,max}}^{T_{\rm m,min}}\frac{2\pi}{d^{\rm 2}}B_{\rm \nu}(T)T^\frac{2-qm}{qm}dT.
\end{eqnarray}
Here $R_{\rm \star}$ and $T_{\rm \star}$ are the radius and effective temperature of the star. The optically thick inner rim and midplane of the disk are assumed to have power-law temperature distributions between the $T_{\rm r}$ and $T_{\rm m}$ boundaries, with exponents denoted by $qr$ and $qm$. The $D_{\rm 0}$, $D_{\rm 1}$, and $D_{\rm 2}$ parameters denote scaling factors for the emission from the stellar photosphere, the hot inner rim, and the disk midplane, respectively.

Our dust model includes amorphous silicates with olivine and pyroxene stoichiometry, crystalline silicates forsterite, enstatite, and amorphous silica. For each amorphous species, we include three grain sizes  with radii of 0.1, 1.5, and 6.0\mum, and for each crystalline species we include only two grain sizes  with radii of 0.1 and 1.5\mum, since there was no evidence for large crystals (6.0\mum) in our data.  The opacity curves for the amorphous species were calculated using MIE theory, and those for the crystalline species were derived using a distribution of hollow spheres (DHS) approximation \citep{2009ApJ...695.1024J}. The optical constants were taken from  \citet{Servoin73} for forsterite, \cite{1995A&A...300..503D} for amorphous silicate with olivine and pyroxene stoichiometry, \cite{1997A&A...327..743H} for silica, and \cite{1998A&A...339..904J} for enstatite. To estimate the uncertainties in the derived abundances, the TLTD routines apply a simple and well-established Monte Carlo technique. The TLTD routines add normally distributed noise to the observed spectra with an amplitude given by the flux uncertainty in each spectral channel, thereby generating 100 versions of each spectrum that are all consistent with the data. For each of these spectra, the TLTD routines then perform the same compositional fit as described above, yielding 100 values for each fit parameter. The mean is then adopted as the best-fit value for each parameter, and the standard deviation in the positive and negative directions yields the 1\,$\sigma$ uncertainty on each parameter. The fit results will be presented and discussed in Sect.~\ref{Sec:dust_for_epsilon_cha}.

\section{Results and discussion}\label{Sec:result} 
\subsection{Stellar properties}
In this section, we will first confirm the memberships of the $\epsilon$\,Cha stars using proper motion data and then estimate the masses and ages of the members.

\subsubsection{Proper motions}
\label{Sec:proper_motion}

\begin{figure}[h]
\centering
\includegraphics[width=0.95\columnwidth]{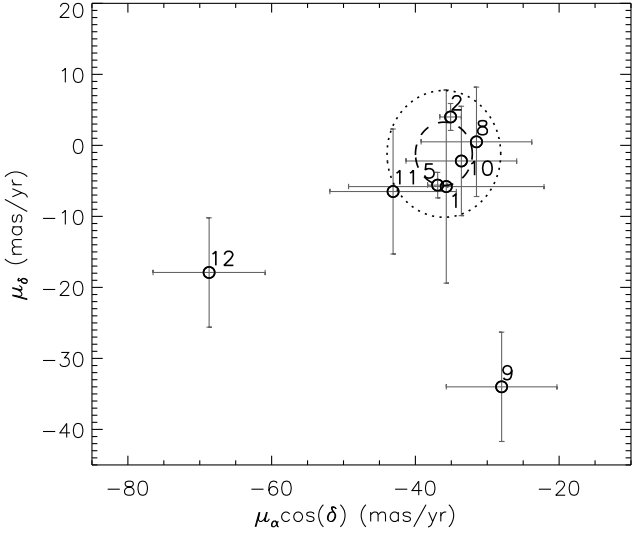}
\caption{Proper motions of known stars in the $\epsilon$\,Cha association taken from the catalog by \cite{2010AJ....139.2440R}. The mean proper motion of the members and the 1 and 2\,$\sigma$ ellipses of the standard deviation of the distribution are shown with dashed-line and dotted-line curves, respectively.}\label{Fig:Proper_motion}
\end{figure}

\begin{figure}[h]
\centering
\includegraphics[width=0.95\columnwidth]{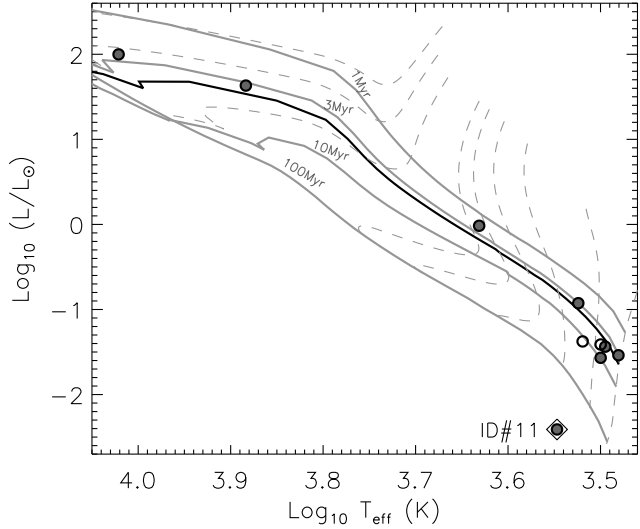}
\caption{HR diagram of members (filled circles) of $\epsilon$\,Cha. The open diamond marks the underluminous object ID\#11. The grey lines are PMS isochrones of 1,  3, 10, and 100\,Myr, and the dark line is the PMS isochrone of 4\,Myr \citep{2008ApJS..178...89D}. The dashed lines are the PMS evolutionary tracks for individual masses. The open circles are for stars ID\#9 and 12, which have proper motions that deviate from the other stars.}\label{Fig:HRD}
\end{figure}

Proper motions provide kinematic information of stars, which is often a very useful discriminant to separate members of a cluster or stellar association from unrelated field objects \citep[see, e.g.,][]{2010ApJ...716L..90R}. Recently, \citet{2010AJ....139.2440R} presented a new catalog of proper motions based on the International Celestial Reference System (ICRS) using a combination of USNO-B1.0 and 2MASS astrometry. This catalog provides proper motions of eight $\epsilon$\,Cha members, which are shown in \fig~\ref{Fig:Proper_motion} and listed in Table~\ref{Tab:acc}. It is clear that most stars have very similar proper motions, with the exception of the stars ID\#9 and 12, which move in different directions. We calculate the weighted mean of the proper motions of the group members, excluding the latter two sources, and find an average proper motion of $\mu_{\alpha}$=$-$36.0\,mas/yr and $\mu_{\delta}$=$-$1.2\,mas/yr for the $\epsilon$\,Cha association, corresponding to a projected linear velocity of $\sim$19.5\,km/s. The 1 and 2\,$\sigma$ error ellipses are also shown in \fig~\ref{Fig:Proper_motion}. Since 
the stars with ID\# 9 and 12 have proper motions that both differ by $\gtrsim$3\,$\sigma$ from the average group value, this casts doubt on their membership to the $\epsilon$\,Cha association. No proper motion estimates are available for the sources with ID\# 3, 4, 6, and 7. Since these sources show X-ray emission and are spatially associated with HD\,104237A, it is highly likely that they are $\epsilon$\,Cha members. In summary, we confirm ten of the previously identified members as  members of  $\epsilon$\,Cha, and challenge the membership of two stars, ID\#9 and 12. The refined memberships of $\epsilon$\,Cha are also listed in Table~\ref{Tab:acc}.  We must stress that we cannot exclude with certainty that sources ID\#9 and 12 are members of the $\epsilon$\,Cha association. Both sources may be in binary systems, where their orbital motions can alter their proper motions, or they may have been ejected from binary systems. Moreover, the  strength of Li absorption lines in these two stars is comparable to other $\epsilon$\,Cha stars,  as are the Na\,I and K\,I strengths, indicating that the two stars have ages  similar to those of the other $\epsilon$\,Cha members \citep{2004ApJ...616.1033L, 2009MNRAS.400L..29L}. Source ID\#9 is also an X-ray emitter \citep{2003ApJ...599.1207F}. Thus both sources are clearly young stars and may well have formed together with the other $\epsilon$\,Cha members.

\subsubsection{The masses and ages of the  $\epsilon$\,Cha members}

Figure~\ref{Fig:HRD} shows the HR~diagram of the sources in the $\epsilon$\,Cha association, where the temperatures were adopted from the observed spectral types \citep{1995ApJS..101..117K,2003ApJ...593.1093L} and the luminosity was determined as described in Sect.~\ref{Sec:targets}. In the figure, the theoretical pre-main sequence (PMS) evolutionary tracks are obtained from \citet{2008ApJS..178...89D}. The masses and ages were then deduced by comparing the location of each object in the HR~diagram with the PMS evolutionary tracks \citep{2008ApJS..178...89D}. For the suspicious members ID\#9 and 12, masses and ages were estimated, assuming that they are at the same distance as the  $\epsilon$\,Cha\, association. {\newnewrev  As a comparison, we also estimated the masses and ages using the  PMS evolutionary tracks from \citet{1998A&A...337..403B}. In general, both PMS  evolutionary tracks give similar masses and ages for most of sources in our sample.}

\begin{figure}[h]
\centering
\includegraphics[width=0.95\columnwidth]{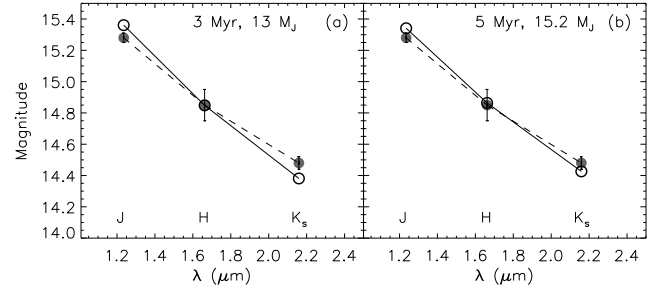}
\caption{Observed photometry (filled circles) and model photometry (open circles) for HD\,104237C in the $J$, $H$, and $K_{\rm s}$ bands. Panel (a) shows model photometry for a 13\,\Mjup\ brown dwarf with an age of 3\,Myr, panel (b) shows a model of a 15.2\,\Mjup\ object of 5\,Myr. The model photometry was adopted from \citet{2000ApJ...542..464C}.
}\label{Fig:BD}
\end{figure}

The resulting ages and masses for all stars are listed in Table~\ref{Tab:acc}. The ages of all the confirmed members, with the exception of CXOU\,J115908.2, are around 3--5\,Myr, which is consistent with the previous estimate \citep{2003ApJ...599.1207F}. The source ID\#11 appears sub-luminous by a factor $\sim$30 with respect to objects of similar spectral type, placing it below the zero-age main sequence (ZAMS) and inhibiting a mass and age estimate through placement in the HR~diagram. We estimated its mass from its spectral type (M2.25) by assuming an age of 4\,Myr. There is no estimate of the spectral type of source HD\,104237B (ID\#4). We estimated its mass from its position in the $J$\,vs.\,$J-H$ color-magnitude diagram, assuming an age of 4\,Myr and using model $J$ and $H$ magnitudes from the PMS evolutionary tracks \citep{1998A&A...337..403B,2008ApJS..178...89D}. The results are listed in Table~\ref{Tab:acc}.

\begin{figure}
\centering
\includegraphics[width=0.95\columnwidth]{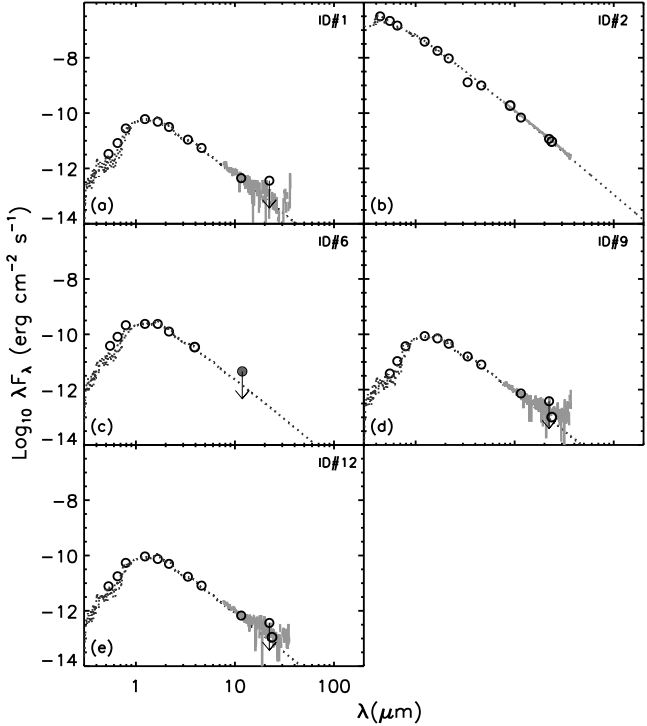}
\caption{SEDs of apparently diskless stars. The photospheric emission level is indicated with a dotted curve in each panel. The circles show the photometry in different bands. The filled circle show the upper limit derived from our VLT/VISIR imaging for ID\#6. The thick grey solid lines in panels a, b, d and e show the Spitzer IRS spectra.}\label{Fig:SED1}
\end{figure}

HD\,104237C, located $\sim$5\farcs3 to the northwest of HD\,104237A, is a source of X-ray emission seen by Chandra \citep{2003ApJ...599.1207F} with a near-infrared counterpart \citep{2004ApJ...608..809G}. At a distance of 114\,pc, the projected distance between HD\,104237C and HD\,104237A is $\sim$604\,AU. Figure~\ref{Fig:BD} shows the observed photometry for HD\,104237C in the $J$, $H$, and $K_{\rm s}$ bands from \cite{2004ApJ...608..809G}. Since no spectral-type estimate is available for HD\,104237C,  we estimate its mass by comparing the observed near-infrared photometry to model photometry for young brown dwarfs at ages of 3 and 5\,Myr  \citep{2000ApJ...542..464C}. The best-fit masses for the assumed ages and the assumed distance of 114\,pc are 13.0 and 15.2 Jupiter masses (\Mjup), respectively. We will adopt these numbers as the mass range for HD\,104237C, putting this object at the boundary between a very low-mass brown dwarf and a ``free-floating planet''. Of the currently known members of the $\epsilon$\,Cha association, HD\,104237C has by far the lowest mass.

\subsection{Disk properties}

In this section, we first estimate the rates at which material is accreted onto the central stars in the $\epsilon$\,Cha member systems and then characterize their disks in terms of evolutionary state. Finally, we derive the dust properties of those protoplanetary disks that show emission from silicate dust in their IRS spectra and compare them with the dust properties found for protoplanetary disks in other sparse stellar associations or star-formation regions.

\subsubsection{Accretion}\label{Sec:Accretion}

{\rev We used the luminosity of the H$\alpha$ emission line as a proxy for the accretion rate. The line luminosity was obtained by integrating over the line profile, adopting the best-fit model atmosphere spectrum (see Sect.~\ref{Sec:targets}) as the continuum level.} We estimated the accretion rates from the observed H$\alpha$ emission-line luminosity using the empirical relation between the latter quantity and the accretion luminosity derived in \citet{2009A&A...504..461F}:
\begin{equation}
log(L_{\rm{acc}}/L_{\odot})=(2.27\pm0.23)+(1.25\pm0.07)\times log (L_{\rm{H}\alpha}/L_{\odot}).
\end{equation}

The inferred accretion luminosities are then converted into mass accretion rates using the following relation:

\begin{equation}
\dot{M}_{\rm acc}=\frac{L_{{\rm acc}}R_{\star}}{{\rm G}M_{\star} (1-\frac{R_{\star}}{R_{\rm in}})},
\end{equation}

\noindent where $R_{\rm in}$ denotes the truncation radius of the disk, which is taken to be 5\,$R_{\star}$ \citep{1998ApJ...492..323G}, G is the gravitational constant, $M_{\star}$ is the stellar mass as estimated from  the location of each star in the HR~diagram, and $R_{\star}$ is the stellar radius derived using the SED fitting procedure described in Sect.~\ref{Sec:targets}. There are four sources in $\epsilon$\,Cha showing signs of active accretion, and their accretion rates are listed in Table~\ref{Tab:acc}.

\begin{figure*}[ht]
\centering
\includegraphics[width=1.6\columnwidth]{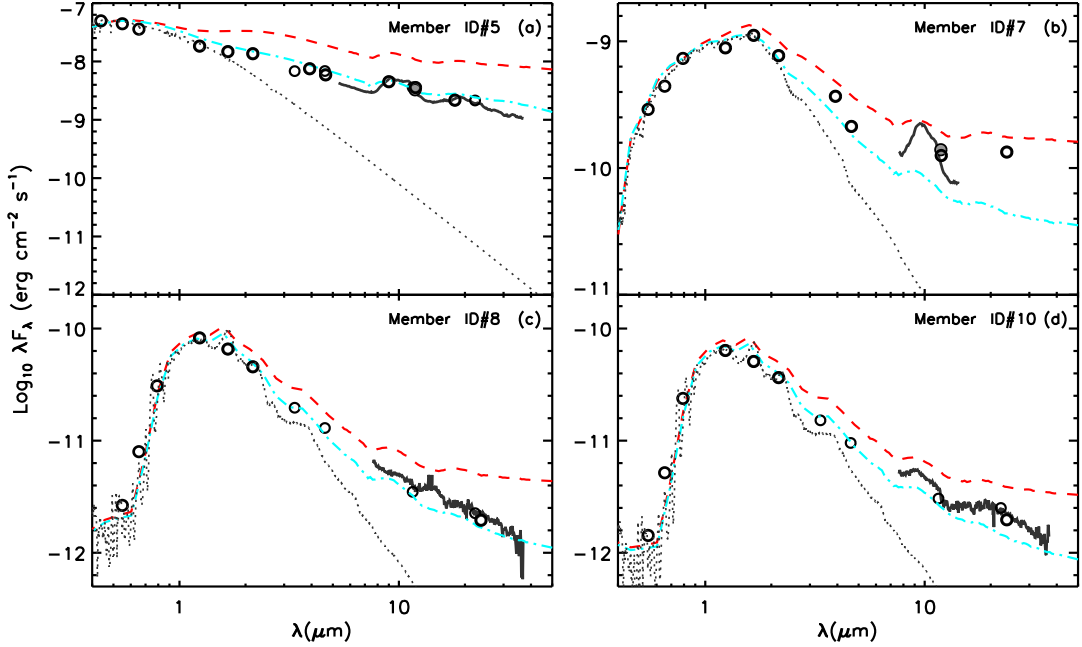}
\caption{SEDs of the disk population in $\epsilon$\,Cha. The photospheric emission level is indicated with a grey dotted curve in each panel. The circles show the photometry in various bands. The filled circles show the photometry from our VLT/VISIR imaging. The solid line shows the IRS spectrum in each panel. The dashed line in each panel represents the calculated SED of a flaring disk model, and the dash-dotted line shows the calculated SED of a flattened disk model.}\label{Fig:SED2}
\end{figure*}

\subsubsection{Demographics of disk population in  $\epsilon$\,Cha}\label{Sec:demographics}

\begin{figure*}
\centering
\includegraphics[width=1.8\columnwidth]{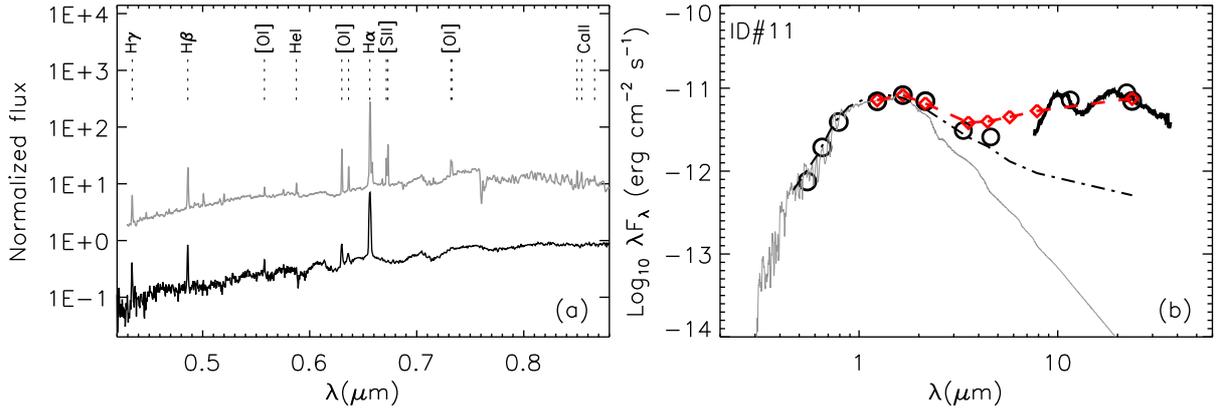}
\caption{Optical spectrum and SED of 2MASS\,J12014343 (ID\#11). (a) The optical spectrum of source ID\#11 (black line) and of a similarly underluminous object in L1641 \citep[L1641-ID\#122, grey line, ][]{2009A&A...504..461F}. (b) The SED of source ID\#11. The photospheric emission level is indicated with a grey solid curve. The open circles show the broad-band photometry of ID\#11, and the IRS spectrum of this source is shown in black. The photometry of underluminous source L1641-ID\#122 is depicted with the open squares connected by dashed line for comparison and shows striking resemblance to that of ID\#11 at mid-infrared wavelengths. The dash-dotted line presents the median SED of the distributed disk population in L1641 \citep{2009A&A...504..461F}.\label{Fig:edge_on}}
\end{figure*}

In Figs.~\ref{Fig:SED1}, \ref{Fig:SED2}, and~\ref{Fig:edge_on}(b), we show the SEDs of the ten sources for which spectral types are available, together with the best-fit stellar model atmospheres (see Sect.~\ref{Sec:targets}). In total, we detected nine sources with the Spitzer IRS. Among these, seven were confirmed as members of the $\epsilon$\,Cha association, based on common proper motion in Sect.\,\ref{Sec:proper_motion}. Two of the members, CXOU\,J115908.2 (ID\#1) and $\epsilon$\,Cha\,AB (ID\#2), do not show significant excess emission over the stellar photosphere in the IRS spectral range. Also, ID\#9 and 12, whose memberships were challenged based on their proper motions, do not show infrared excess emission in the IRS spectra, strongly suggesting that they do not (any longer) harbor circumstellar disks. For the sources that were not detected with the IRS, we characterized their disks based on infrared photometric data only. The source HD\,104237D (ID\#6) cannot be spatially separated from HD\,104237A with the IRS and MIPS instruments due to the comparatively poor spatial resolution of the Spitzer Space Telescope at long wavelengths. HD\,104237D is not detected in our VISIR imaging at 12\mum, and we derived a 3$\sigma$ upper limit of $\sim$18\,mJy for its flux at this wavelength. This upper limit is just above the expected photospheric emission level, as extrapolated from shorter wavelengths and shown in \fig~\ref{Fig:SED1}, indicating that HD\,104237D no longer has a disk. However, based on currently available data, the possibility of an evolved disk with a large inner hole that would be observable only at longer wavelengths cannot be excluded. The underluminous object 2MASS\,J12014343 (ID\#11) shows a very flat SED, which is similar to the SEDs of similarly underluminous objects discovered in other star-forming regions \citep[see][]{2009A&A...504..461F}. For sources HD\,104237B (ID\#4) and HD\,104237C (ID\#3), we only have photometry in $JHK_{\rm s}$ bands. The $J-H$ vs. $H-K_{\rm s}$ color-color diagram shows only HD\,104237B to still have near-infrared excess emission, indicative of a hot, optically thick inner disk \citep{2004ApJ...608..809G}.

In \fig~\ref{Fig:SED2}(a), we show the SED of HD\,104237A, which exhibits strong excess emission at wavelengths beyond $\sim$2\mum. HD\,104237A therefore shows clear evidence for a hot, optically thick inner disk. {\newnewrev We use the 2D RADMC code from \citet{2004A&A...417..159D} to calculate two  SEDs, one for a flaring disk model and the other for a flattened disk model, given the effective temperature ($T_{\rm eff}$), stellar radius ($R_{\star}$), and stellar mass ($M_{\star}$) of HD\,104237A. In both models, we assume a pressure scale height ($H_{\rm P}$) that varies as a power law with the radius (R), $H_{\rm P}/R=R^{1/7}$. The inner disk radius  ($R_{\rm in}$)  is set at the dust destruction radius of $T_{\rm dust}\sim$1500\,K, and the outer disk radius ($R_{\rm out}$) is 500\,AU. The scale height at the outer disk radius  is set as  $H_{\rm out}$/$R_{\rm out}$=0.2 for the flaring disk model and as $H_{\rm out}$/$R_{\rm out}$=0.1 for the settled/flattened disk model. We use a power-law size distribution with an exponent $-3.5$ for dust grain with a minimum grain size of 0.1\mum\ and a maximum grain size of 1000\mum. Two populations of amorphous dust grains (25\% carbon and 75\% silicate) have been included in the models. In the flaring disk model, the disk mass is set to be 0.02\,$M_{\star}$, and in the flattened disk model, the disk mass ($M_{\rm disk}$) is 0.001\,$M_{\star}$. The gas-to-dust ratio is taken to be 100 in the models. The disk surface density ($\Sigma$) is estimated from the total disk mass, assuming $\Sigma\propto R^{-1}$. In \fig~\ref{Fig:SED2}(a), we show the two model SEDs for the disk inclination angle $\sim55^{\circ}$. The SED of  HD\,104237A is more similar to the SED for the flattened disk model than to the SED for a flaring disk model.}

In Fig.~\ref{Fig:SED2}(b), (c), and (d), we show the SEDs of group members HD\,104237E (ID\#7), USNO-B120144.7 (ID\#8), and 2MASS\,J12005517 (ID\#10), {\newnewrev as well as two model SEDs: one for a flaring disk and one for a flattened disk model for each object. In the calculation, we use  the same parameters in the disk models as those for HD\,104237A,  except that the outer disk radius is set to be 200\,AU.  The actual stellar parameters of each object are used in the calculation. Overall, the SEDs of the three objects  HD\,104237E (ID\#7), USNO-B120144.7 (ID\#8), and 2MASS\,J12005517 (ID\#10) are between the two model SEDs. It can also be seen that the excess emission in the MIPS 24\mum\ band of source ID\#7 is similar to the model SED of the flaring disk, whereas its infrared excess at shorter wavelengths is substantially lower and  closer to the SED of a flattened disk. This may indicate that source ID\#7 has started to dissipate its inner disk, while its outer disk remains relatively intact. In contrast to object ID\#7, the very low-mass objects ID\#8 and 10 show a fairly uniform depletion of the SED. Their SEDs are similar to the SEDs of the flattened disk models.}

Spitzer observations of clusters with ages of several Myrs, such as IC\,348, NGC\,2362, $\eta$\,Cha, and the Coronet cluster,  suggest that, qualitatively, two evolutionary paths exist for going from a primordial to a debris disk configuration {\newnewrev \citep{2006AJ....131.1574L,2008ApJ...687.1145S,2009AJ....138..703C,2009ApJ...698....1C,2009ApJ...701.1188S,2011ApJ...732...24C}}. Each path shows a characteristic behavior of the SED: (1) some objects show little or no excess emission in the shorter IRAC bands and strong excess emission at 24\mum, suggesting that disks are dissipated in an inside-out fashion; (2) some objects show infrared excess of an approximately uniformly reduced magnitude compared to primordial disks over all wavelengths out to 24\mum, suggesting a reduction of the effective disk height, i.e., the height above the disk midplane where the disk becomes optically thick to the stellar radiation. A reduced disk height causes a smaller fraction of the stellar energy to be absorbed and reprocessed by the disk and thereby yields an infrared excess of reduced magnitude. This may occur if dust coagulation takes place in the disk, causing the grains to couple somewhat less well to the gas and allowing them to settle towards the midplane. How effective this is  also depends on the level of turbulence in the disk; grains settle more easily in a disk with low turbulence. {\newnewrev In some cases, the disks could have lost a large number of small dust grains, and become optically thin. This type of disk can also show a significant depletion of infrared emission \citep{2011ApJ...732...24C}.} In $\epsilon$\,Cha, source ID\#7 shows an SED reminiscent of the first type of disk evolution (inside-out fashion), whereas sources ID\#8 and 10 show an SED corresponding to the second scenario (globally depleted). Both latter sources show strong evidence of dust growth, which seems to have progressed particularly far in source ID\#8 judging from the absence of silicate features at $\sim$10 and 20\mum. The absence indicates that there is no substantial population of dust grains with sizes of $\lesssim$10\mum\ left in the atmosphere of this disk. It is interesting to note that source ID\#7, which shows evidence of an inside-out disk clearing, has a silicate feature emitted by predominantly small, sub-micron sized grains.

\subsubsection{2MASS J12014343: a disk system with high inclination?}
\label{Sec:edg_on}

When the HR diagram of the $\epsilon$\,Cha sample (\fig~\ref{Fig:HRD}) is inspected, one object stands out in the sense that it lies well \emph{below} the ZAMS. Compared to $\sim$4\,Myr old stars with the same spectral type, 2MASS\,J12014343 (ID\#11) appears underluminous by a factor of $\sim$30. Since evidence of group membership is strong (optical emission lines, infrared excess, and in particular a proper motion in agreement with the $\epsilon$\,Cha group, see Sect.~\ref{Sec:proper_motion}) its underluminosity cannot be simply explained due to ID\#11 being a field star. To understand its nature we should resort to other explanations. In \fig~\ref{Fig:edge_on}(a), we show the optical spectrum of  ID\#11, on which  Balmer emission lines and forbidden oxygen emission lines with very high EWs are seen. Source ID\#11 also shows a particularly strong infrared excess (\fig~\ref{Fig:edge_on}(b)). Objects with these characteristics have been found in other star-forming regions, e.g., Lupus 3 dark cloud, Taurus, L1630, and L1641 \citep{2003A&A...406.1001C,2004ApJ...616..998W,2009A&A...504..461F,2010ApJ...718.1200M}. One possible explanation for this phenomenon is that these sources are systems harboring flared disks with moderately high inclinations where the stellar photospheric light is largely absorbed by the material in the cold, flared outer disk. We still receive photospheric light, but a large fraction of the light we see is scattered off the disk surface and has a much reduced total flux. The optical emission lines of large EWs may arise in an outflow or disk wind. They need not intrinsically be brighter than in similar objects with ``normal'' apparent luminosities; it is the reduced strength of the continuum flux, not the intrinsic line strength, that causes the \emph{EWs} to be high. This scenario only works if the line-forming region is much larger than the central star, so that at least part of the line flux reaches us relatively unhindered, while the photospheric continuum is strongly absorbed. In particular, this scenario also explains why some of these apparently underluminous objects show some emission lines like \HeI\,5876\AA\ and the \CaII\, near-infrared triplet (8498, 8542, and 8662~\AA) with EWs that appear \emph{not} enhanced \citep[see][]{2003A&A...406.1001C,2009A&A...504..461F}: these lines are mainly formed in the magnetospheric infall flows \citep{1998ApJ...492..743M}, which are close to the stellar surface and should be as much occulted as the photospheric continuum. {\rev \citet{2004ApJ...616.1033L} suggested that 2MASS~J12014343 harbors an edge-on {\newnewrev disk} and seen in scattered light. The 10\mum\ silicate feature of 2MASS J12014343 is seen in emission, suggesting that its inclination is substantially different from 90 degrees (edge on).}

In \fig~\ref{Fig:edge_on}, we also compare the optical spectrum and SED of source ID\#11 with an object that shows very similar characteristics and was recently found in the Lynds~1641 cloud in Orion \citep[ID\#122, named L1641\#122 hereafter, see][]{2009A&A...504..461F}. L1641\#122 is also apparently underluminous compared to stars of similar spectral type in the same cloud by a factor of $\sim$30, similar to ID\#11. But contrary to ID\#11, it shows the \HeI\,5876\AA, \SII, and \CaII\, near-infrared triplet (8498, 8542, and 8662~\AA) in emission, indicating active accretion. The optical spectrum of ID\#11 did not show signs of active accretion at the time of observation. In \fig~\ref{Fig:edge_on}(b), we compare the median SED of the distributed disk population in L1641 (L1641D) and that of L1641\#122 \citep{2009A&A...504..461F} with the SED of ID\#11. From this comparison we can see that the infrared SEDs of ID\#11 and  L1641\#122 are remarkably similar and that both underluminous objects show much stronger mid-infrared excesses than the average source in L1641, even though the L1641D population is much younger ($\sim$1\,Myr) than $\epsilon$\,Cha.  At wavelengths shorter than $\sim$8\mum, the SED of ID\#11 shows a rapidly decreasing trend. This is different from the SED of L1641\#122, which shows relatively flat SED till near-infrared wavelengths. It suggests that the inner disk around ID\#11 may have been dissipated, consistent with its very low or no accretion activities.

\begin{figure}
\centering
\includegraphics[width=1.\columnwidth]{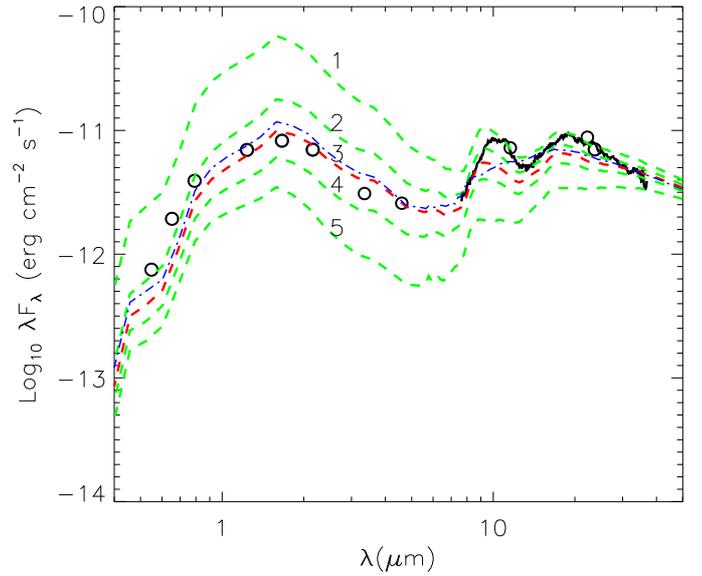}
\caption{{\newnewrev SED of the source ID\#11. The open circles show the broad-band photometry of ID\#11, and the IRS spectrum of this source is shown in black. The dashed lines marked by numbers 1-5 show model SEDs for a same model disk with different inclination angles. From model 1 to model 5,  the inclination angles are 83$^{\circ}$, 84$^{\circ}$, 84.5$^{\circ}$, 85$^{\circ}$, and 86$^{\circ}$, respectively. In the model, we added a few  small dust grains to the hole of the disk. The model with disk inclination $\sim$84.5$^{\circ}$ best fits the SED of ID\#11. The dash-dotted line shows the model SED for a disk (inclination angle$\sim$84$^{\circ}$) without small dust grains in the disk hole. The model SED generally fits the shape of the observed SED of ID\#11, but fails to reproduce the observed silicate features.} \label{Fig:fit_edge_on}}
\end{figure}

{\newnewrev As discussed above, based on the SED of ID\#11, the circumstellar disk of ID\#11 may have two properties: (1) the inclination angle of the disk should be high but less than edge on, (2) the inner region of the disk may be evolving. To explore the possibility that such types of disks can produce SEDs similar to the one of ID\#11, we performed a detailed SED modeling using the 2D RADMC code \citep{2004A&A...417..159D}. In the calculation, the stellar radius and mass for ID\#11 is given as a 4\,Myr PMS star with a spectral type of M2.25, which is  $\sim$1\,$R_{\odot}$ and $\sim$0.3\,$M_{\odot}$ from \citet{2008ApJS..178...89D}. By varying $M_{\rm disk}$, $R_{\rm out}$, $R_{\rm in}$, and disk inclination angle, we found that the shape of the SED of ID\#11 can be generally fitted by the SED of a disk with  $R_{\rm in}\sim$0.5\,AU, $M_{\rm disk}\sim$0.0001\,$M_{\star}$, the inclination angle $\sim$84$^{\circ}$, and  $H_{out}/R_{\rm out}\sim$0.1 (see Fig.~\ref{Fig:fit_edge_on}). The dust populations in the disk are same as those in the disk models described in Sect.~\ref{Sec:demographics}. However, the SED of such a disk model fails to reproduce the observed 10 and 20\mum\ silicate features for ID\#11. In order to produce these silicate features, we added a few small dust grains in the disk hole. The dust grains are located close to the inner rim of the disk. The grain size distribution is a power-law distribution with an exponent $-3.5$ for dust grain  with a minimum grain size 0.005\mum\ and a maximum grain size 0.5\mum. We fixed the  $M_{\rm disk}\sim$0.0001\,$M_{\star}$ and $H_{\rm out}/R_{\rm out}\sim$0.1. By varying the $R_{\rm in}$, inclination angle, the mass of small dust grains, and the location of these small dust grains, we found that the SED of ID\#11 can be fitted when $R_{\rm in}\sim$0.7\,AU, the inclination angle $\sim$84.5$^{\circ}$, and the total mass of the small grains $\sim10^{-11}$\,\Msun, which are located between 0.6 and 0.7\,AU from the central star.}

\subsubsection{The disk and accretor frequencies in $\epsilon$\,Cha}
\label{Sec:disk_frequency_eps_cha}

In order to estimate the disk frequency in  $\epsilon$\,Cha, i.e., the fraction of sources that shows evidence of a circumstellar disk in the form of infrared excess emission, one needs to identify all members of the  $\epsilon$\,Cha association and characterize their infrared SEDs. This is a challenging task because the $\epsilon$\,Cha association is spread over a large area of the sky and a full inventory is beyond the scope of the current work. Instead, we restrict ourselves to the region within a radius of 0\fdg5 of $\epsilon$\,Cha\,AB. This part of the sky has been surveyed by \citet{2004ApJ...616.1033L}, and our knowledge of the association can be considered  essentially complete. In Figs.~\ref{Fig:SED1}, ~\ref{Fig:SED2}, and ~\ref{Fig:edge_on}(b), we show the SEDs of ten members with published spectral types. Among these, five objects show excess emission at near- or mid-infrared wavelengths and thus are harboring circumstellar disks. We also include HD\,104237B, which shows excess emission at $K_{\rm S}$ bands according to \cite{2004ApJ...608..809G}, and obtain a disk frequency among the 11 stellar members of $\epsilon$\,Cha association of {\newnewrev 55$_{-15}^{+13}\%$} (6/11)\footnote{The uncertainties in disk frequencies for small samples are all estimated in the manner described by \citet{2003ApJ...586..512B}.}. This is quite high for a 3--5\,Myr old population, though this value is clearly subject to low-number statistics and it is premature to conclude that the disk frequency in $\epsilon$\,Cha is high based on these data alone. However, as we will see later in this section, the five other sparse associations of which we estimated the disk frequencies, Taurus, the Coronet cluster (CrA), MBM\,12, $\eta$\,Cha, and TW\,Hya, show a similar trend. The ensemble of data provides strong evidence that disk lifetimes in sparse associations are longer than those in more crowded environments.

\begin{figure}[h]
\centering
\includegraphics[width=0.95\columnwidth]{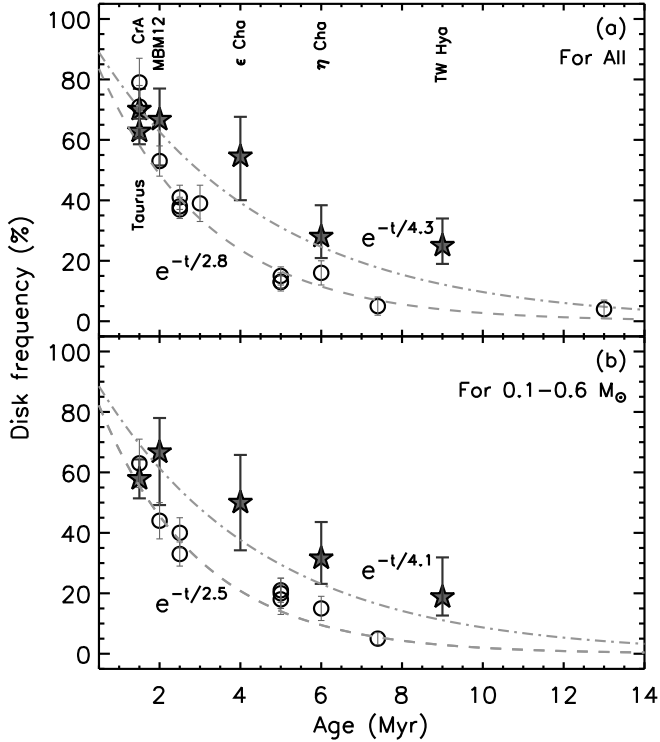}
\caption{Disk frequencies for different clusters/star-formation regions plotted as a function of their ages, which are all estimated from \citet{1998A&A...337..403B} (see Table~\ref{Tab:disk_frequency}).  In panel (a) we plot the disk frequency among all known members, in panel (b) we show only the low-mass (0.1--0.6\,\Msun) population. The filled pentagram represents the sparse stellar associations, Taurus, CrA, MBM\,12, $\epsilon$\,Cha, $\eta$\,Cha, and TW~Hya in panel (a), and  the sparse stellar associations, Taurus, MBM\,12, $\epsilon$\,Cha, $\eta$\,Cha, and TW~Hya in panel (b). The open circles show the disk frequencies in the compact clusters or the OB associations (see Appendix~\ref{Appen:disk_fractions} for a detailed description). In each panel, the dashed line represents the best-fit exponential decay for all regions, excluding the sparse associations and the dash-dotted line represents the best-fit exponential decay for the sparse associations.}\label{Fig:df}
\end{figure}

In \fig~\ref{Fig:df}, we illustrate this by showing the estimated disk frequencies in the aforementioned sparse associations and a number of other star-formation regions as a function of their age (see Appendix~\ref{Appen:disk_fractions} for a detailed description). {\newnewrev The disk frequencies are estimated using only IRAC data or infrared data at wavelengths shorter than $\sim$8\mum, and the ages are all estimated from \citet{1998A&A...337..403B}}. Figure~\ref{Fig:df} also shows a fit to the observed disk frequencies of all regions, \emph{except} the six sparse associations, of the form $f_{\rm disk}$=exp($-t/\tau_0$). We find a value of {\newnewrev $\tau_0$\,$=$\,2.8$\pm$0.1\,Myr} to yield a good fit {\newnewrev (reduced $\chi^2$=1.8)} to the overall distribution. This agrees  with the value of $\sim$3.0\,Myr found by  \citet{2010A&A...510A..72F} in an earlier, similar study. Because the disk lifetime in any given environment may depend on the mass of the central star \citep[e.g.,][]{2009ApJ...695.1210K} it is useful to investigate the low-mass population separately. In \fig~\ref{Fig:df}(b), we show the disk frequency among stars with estimated masses in the 0.1$-$0.6\,\Msun \ range for the same star-forming regions as shown in \fig~\ref{Fig:df}, but including only those for which the low-mass population has been characterized. We find essentially the same typical disk dispersal time as for the whole mass range {\newnewrev($\tau_0$\,$=$\,2.5$\pm$0.2\,Myr)}  {\newnewrev (reduced $\chi^2$=1.1)}. Again, the sparse associations are deviant in the sense that they show systematically higher disk frequencies.  {\newnewrev We fitted the disk frequencies in the sparse associations  of the form $f_{\rm disk}$=exp($-t/\tau_0$), which gives  $\tau_0$\,$=$\,4.3$\pm$0.3\,Myr  {\newnewrev (reduced $\chi^2$=1.6)} among all sources and $\tau_0$\,$=$\,4.1$\pm$0.5\,Myr   {\newnewrev (reduced $\chi^2$=1.5)} among sources with masses in the 0.1$-$0.6\,\Msun \ range. The estimated lifetimes of disks in  the sparse associations are  longer than those in the dense (cluster) environments with more than 2$\sigma$ confidence.} As noted in Fig.~\ref{Fig:df}, disk frequencies for the sparse associations are nearly constant at 60--70\% for ages  below 4\,Myr, and decrease slowly at higher ages. This is suggestive of a bimodal mechanism responsible for disk dissipation in these regions. Very early in the evolution of the associations, 30--40\% of the disks may have been dissipated  quickly by some efficient mechanism, for example, interaction with binary companions in close systems \citep{2006ApJ...653L..57B,2012ApJ...745...19K}. After the fast disk-dissipation phase, the remaining disks would evolve slowly via other mechanisms, e.g., viscous evolution, planet formation, or photoevaporation.

To estimate the accretor frequency, i.e., the fraction of members showing signs of active accretion, we turn to the H$\alpha$ line as a diagnostic. Estimates of the H$\alpha$ EWs are available for eight of the $\epsilon$\,Cha members, of which  five are classified as accretors according to the criteria in \citet{2009A&A...504..461F}. This brings our estimate of the accretor frequency in  $\epsilon$\,Cha to {\newnewrev 63$_{-18}^{+13}\%$}, again obviously subject to low-number statistics. \citet{2010A&A...510A..72F} investigated the age dependency of accretor frequency and derive an empirical relation $f_{\rm acc}$=$e^{-t/2.3}$, where $t$ is in units of Myrs, by fitting an exponential profile to the observed accretor frequencies of a number of star-formation regions. For $\epsilon$\,Cha, aged 3--5\,Myr, the empirical relation predicts an accretor frequency of {\rev $\sim$10$-$30\%}, substantially below the observed value.

In conclusion, we can state that the disks around stars in sparse associations evolve more slowly than those in denser environments.

\begin{figure}
\centering
\includegraphics[width=0.95\columnwidth]{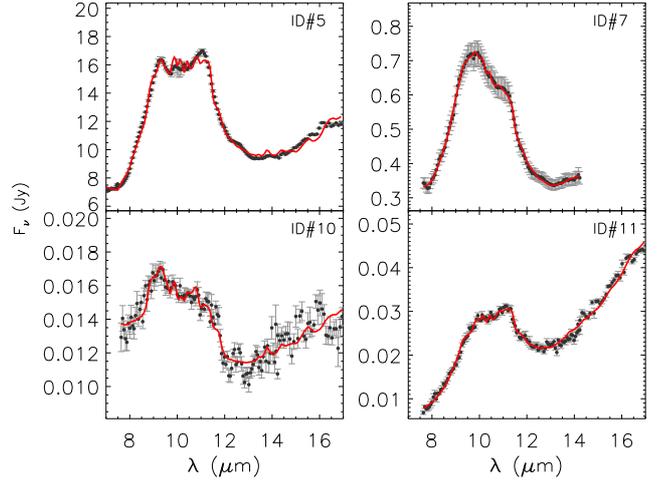} 
\caption{The 10\mum\ silicate features fitted using the TLTD method (solid red lines). The observed spectra are represented as filled circles with the errors in grey.}\label{Fig:fit_IRS_short}
\end{figure}

\begin{figure}
\centering
\includegraphics[width=0.95\columnwidth]{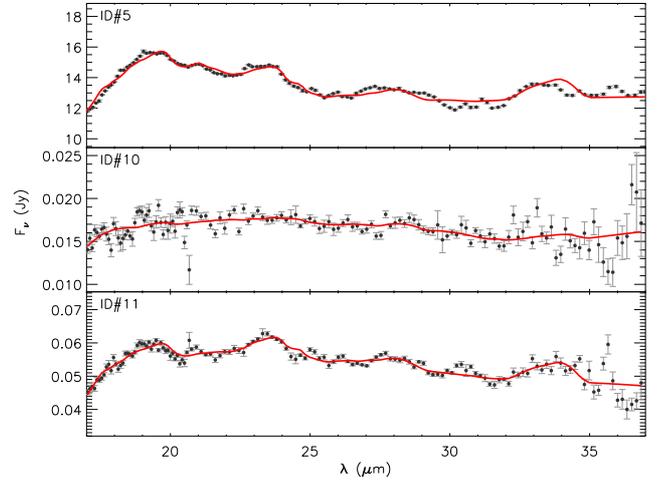} 
\caption{IRS spectra in the 17--37\mum\ range and spectral fits using the TLTD method. The observed spectra are represented as filled circles with the errors in grey, whereas the fits are shown in red.}\label{Fig:fit_IRS_long}
\end{figure}

\subsubsection{The dust properties of disks}\label{Sec:dust_for_epsilon_cha}

We have used the TLTD spectral decomposition method developed by \cite{2009ApJ...695.1024J} to analyze our IRS spectra of the $\epsilon$\,Cha members. In this section, we will first present the derived dust properties of the protoplanetary disks in $\epsilon$\,Cha. Then we will join the $\epsilon$\,Cha sample with a collection of cool T~Tauri stars in the MBM\,12 and $\eta$\,Cha associations, as well as with the Coronet cluster and the sample in the cores-to-disks (c2d) legacy program, for which IRS spectra have been previously analyzed using identical or similar methods. The combined data set is then used to do a statistical study on the dust properties of protoplanetary disks around the cool stars.

\renewcommand{\tabcolsep}{0.12cm}
\begin{table*}
\caption{Silicate compositions estimated from IRS spectra with TLTD. The values are percentages in mass fraction.\label{Tab:dust_for_echa_com}}
\scriptsize
\centering
\begin{tabular}{lcccccccccccccccccccccccccccccc}
\hline\hline
 &  & &\multicolumn{3}{c}{Am. (Olivine-type)}& &\multicolumn{3}{c}{Am. ( Pyroxene-type)}&  &\multicolumn{2}{c}{Forsterite}& &\multicolumn{2}{c}{Enstatite}& &\multicolumn{3}{c}{Silica}\\
\cline{4-6}\cline{8-10}\cline{12-13}\cline{15-16}\cline{18-20}
ID&$\chi^2$ & & 0.1\mum &1.5\mum & 6.0\mum & &0.1\mum &1.5\mum & 6.0\mum & &0.1\mum &1.5\mum & &0.1\mum &1.5\mum & &0.1\mum &1.5\mum &6.0\mum\\
\hline
\multicolumn{20}{c}{Warm disk region}\\
\hline
5& 139.9& & ...&  0.2$^{+0.7}_{-0.2}$&  29.4$^{+1.5}_{-1.8}$& & ...&  17.7$^{+1.2}_{-1.5}$&  40.8$^{+1.9}_{-1.9}$& & 2.8$^{+0.2}_{-0.2}$&  0.1$^{+0.1}_{-0.1}$& & 0.1$^{+0.2}_{-0.1}$&  6.5$^{+0.6}_{-0.6}$& & 1.2$^{+0.1}_{-0.1}$&  0.1$^{+0.1}_{-0.1}$&  1.2$^{+1.0}_{-0.9}$\\
7& 4.8& & 1.7$^{+2.2}_{-1.3}$&  50.2$^{+3.3}_{-3.8}$&  ...& & 0.1$^{+0.7}_{-0.1}$&  21.4$^{+3.0}_{-3.2}$&  1.3$^{+10.1}_{-1.3}$& & 4.8$^{+0.4}_{-0.4}$&  0.1$^{+0.2}_{-0.1}$& & ...&  6.6$^{+1.1}_{-1.3}$& & ...&  0.1$^{+0.1}_{-0.1}$&  14.1$^{+1.9}_{-2.2}$\\
10& 3.8& & ...&  ...&  0.3$^{+1.4}_{-0.3}$& & 33.3$^{+8.1}_{-9.5}$&  11.8$^{+13.7}_{-10.3}$&  0.1$^{+1.3}_{-0.1}$& & 6.5$^{+1.3}_{-1.3}$&  0.1$^{+2.9}_{-0.1}$& & 0.5$^{+1.7}_{-0.5}$&  42.0$^{+4.7}_{-4.9}$& & 5.3$^{+1.1}_{-1.0}$&  0.1$^{+1.9}_{-0.1}$&  ...\\
11& 4.8& & 1.7$^{+2.2}_{-1.3}$&  50.2$^{+3.3}_{-3.8}$&  ...& & 0.1$^{+0.7}_{-0.1}$&  21.4$^{+3.0}_{-3.2}$&  1.3$^{+10.1}_{-1.3}$& & 4.8$^{+0.4}_{-0.4}$&  0.1$^{+0.2}_{-0.1}$& & ...&  6.6$^{+1.1}_{-1.3}$& & ...&  0.1$^{+0.1}_{-0.1}$&  14.1$^{+1.9}_{-2.2}$\\
\hline
\multicolumn{20}{c}{Cool disk region}\\
\hline
5& 57.0& & ...&  17.3$^{+1.3}_{-1.6}$&  1.3$^{+2.9}_{-1.1}$& & ...&  68.3$^{+1.3}_{-1.9}$&  ...& & 4.8$^{+0.1}_{-0.1}$&  0.1$^{+0.1}_{-0.1}$& & 2.0$^{+0.1}_{-0.1}$&  ...& & ...&  ...&  6.3$^{+0.2}_{-0.3}$\\
10& 4.3& & 0.4$^{+9.9}_{-0.4}$&  75.7$^{+8.4}_{-10.7}$&  2.6$^{+19.0}_{-2.6}$& & 0.2$^{+11.1}_{-0.2}$&  6.8$^{+8.4}_{-5.8}$&  0.6$^{+29.9}_{-0.6}$& & 1.2$^{+1.6}_{-0.8}$&  0.1$^{+3.0}_{-0.1}$& & 1.6$^{+1.5}_{-0.9}$&  1.2$^{+3.0}_{-1.0}$& & 0.1$^{+1.0}_{-0.1}$&  0.1$^{+0.8}_{-0.1}$&  9.4$^{+5.3}_{-3.7}$\\
11& 6.3& & ...&  78.6$^{+3.3}_{-2.9}$&  ...& & 10.9$^{+6.2}_{-9.3}$&  4.7$^{+7.5}_{-4.6}$&  0.1$^{+2.1}_{-0.1}$& & 3.2$^{+0.2}_{-0.2}$&  0.1$^{+0.1}_{-0.1}$& & 0.2$^{+0.8}_{-0.2}$&  1.9$^{+0.4}_{-0.6}$& & 0.1$^{+0.2}_{-0.1}$&  0.1$^{+0.1}_{-0.1}$&  0.3$^{+0.7}_{-0.3}$\\
\hline
\end{tabular}
\end{table*}
\normalsize

\vspace{0.3cm}
\noindent
\noindent \textbf{(a) The IRS spectra of the $\epsilon$\,Cha disks} \\
\vspace{-0.3cm}

\noindent
There are five $\epsilon$\,Cha members for which we have IRS spectra that show evidence for a protoplanetary disk. Four of these show the well-known silicate feature between 8 and 13\mum \ in emission (ID\#5,7,10,11, see \fig~\ref{Fig:fit_IRS_short}). Before using TLTD routines to derive the dust composition, we divided the IRS spectra into a ``short'' (7--17\mum) and a ``long'' (17--37\mum) wavelength regime, which are analyzed independently. Since the short wavelength part of the spectrum is dominated by warm disk regions closer to the central star than the cooler regions that dominate long wavelength data, we can search for radial gradients in the dust properties.

\noindent
The derived dust properties are listed in Tables~\ref{Tab:dust_for_echa_com} and ~\ref{Tab:dust_for_echa}. In Table~\ref{Tab:dust_for_echa_com}, we present the outcome of the fits in terms of mass abundances of the different species. In Table~\ref{Tab:dust_for_echa}, we give the mass-averaged grain sizes of the amorphous and crystalline silicates, as well as the fractional contribution of the crystalline species to the total dust mass present in the optically thin disk atmosphere.  As shown in Table~\ref{Tab:dust_for_echa}, the crystalline fractions of the objects in $\epsilon$\,Cha are comparable to those of young stars with similar spectral type in the literature \citep{2008ApJ...687.1145S,2009ApJ...701.1188S,2009A&A...497..379M,2010ApJ...721..431J}. Also, in the warm disk regions, the dust grains of our objects show typical larger sizes than  those in the interstellar medium, suggesting dust growth in these regions. {\rev Based on simulations in a 1D vertical column of a protoplanetary disk, \citet{2011A&A...534A..73Z} suggest that high values of turbulence ($\alpha\sim$0.01) are needed to explain the existence of grains with sizes of a few microns in disk atmospheres at ages of several Myr.}

\noindent
We now briefly describe the individual objects:

\noindent
{\bf HD\,104237A} (ID\#5) \citet{2010ApJ...721..431J} fitted the IRS spectrum of this object using amorphous dust with olivine and pyroxene stoichiometries, as well as silica with grain sizes of 0.1, 2.0, and 5.0\mum, crystalline species forsterite and enstatite with grain sizes of 0.1 and 2.0\mum, and polycyclic aromatic hydrocarbons (PAHs). They divided the IRS spectrum into two sections: 5--17\mum\ and 17--35\mum, which are similar to those we use. From the short wavelength part, they derived mass-averaged grain sizes of the amorphous and crystalline silicates, which are 4.6\mum\ and 1.4\mum, respectively, and a mass fraction of 9.5\% in crystalline silicates. In the long wavelength part of the spectrum, they find 0.1\mum\ and 0.4\mum\ for the mass-averaged grain sizes and 7.1\% for the crystallinity. Even though we use a slightly different set of grain sizes in our fit, our results agree well with those derived by \citet{2010ApJ...721..431J}, with the exception of the mass-averaged grain size of the amorphous silicates derived from the long wavelength channel. Here we find a somewhat higher value. This may be due to the fact that the mass absorption coefficients of amorphous silicates with sizes of 0.1 and 1.5\mum\ are very similar in the 17--37\mum \ spectral range.

\begin{figure}
\centering
\includegraphics[width=0.95\columnwidth]{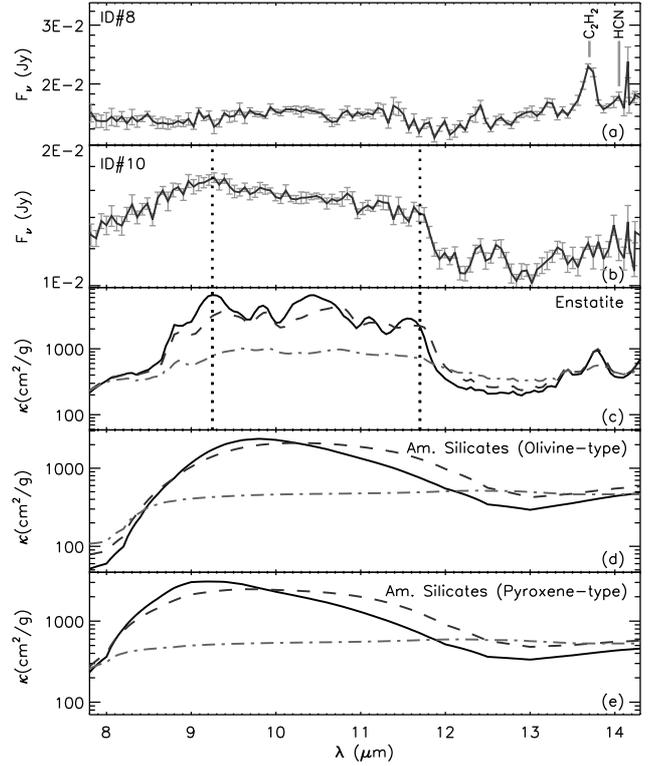}
\caption{(a) IRS spectrum of source ID\#8. The positions of the C$_{2}$H$_{2}$ and HCN rovibrational bands are marked. (b) IRS spectrum of the source ID\#10. The dotted vertical lines mark the features of enstatite. (c), (d), and (e): The mass absorption coefficients of enstatite and amorphous silicates (with olivine and pyroxene stoichiometry) with grain sizes of 0.1\mum (solid lines), 1.5\mum (dashed lines), and 6.0\mum (dash-dotted lines). The dotted lines in panels (b) (c) are identical.}\label{Fig:dust_growth}
\end{figure}

\begin{table}
\caption{Fitting results for IRS sepctra with TLTD. Columns~2, 3, 4: the average sizes of amorphous ($\langle$a$\rangle$$_{\rm am.sil.}^{\rm W}$) and crystalline grains ($\langle$a$\rangle$$_{\rm cryst.sil.}^{\rm W}$) and the mass fractions of crystalline grains (f$_{\rm cryst}^{\rm W}$) are derived from 7-17\mum. Columns~5, 6, 7: the average sizes of amorphous ($\langle$a$\rangle$$_{\rm am.sil.}^{\rm C}$) and crystalline grains ($\langle$a$\rangle$$_{\rm cryst.sil.}^{\rm C}$) and the mass fractions of crystalline grains (f$_{\rm cryst}^{\rm C}$) are derived from 17-35\mum\ \label{Tab:dust_for_echa}}
\centering
\begin{tabular}{lccccccccccccccccccc}
\hline\hline
(1) & (2) & (3) &(4) &(5) &(6) &(7) \\
   &$\langle$a$\rangle$$_{\rm am.sil.}^{\rm W}$&$\langle$a$\rangle$$_{\rm cryst.sil.}^{\rm W}$ &f$_{\rm cryst}^{\rm W}$  &$\langle$a$\rangle$$_{\rm am.sil.}^{\rm C}$&$\langle$a$\rangle$$_{\rm cryst.sil.}^{\rm C}$ &f$_{\rm cryst}^{\rm C}$\\
ID & (\mum) &(\mum) &(\%) & (\mum) &(\mum) &(\%)\\
\hline
5 &5.1$^{+0.1}_{-0.1}$& 1.1$^{+0.1}_{-0.1}$&  9.4$^{+0.9}_{-0.7}$&  1.6$^{+0.2}_{-0.1}$& 0.1$^{+0.1}_{-0.1}$&  6.8$^{+0.2}_{-0.2}$\\
7 &1.5$^{+0.6}_{-0.1}$& 0.9$^{+0.1}_{-0.1}$&  11.4$^{+1.3}_{-1.5}$&  ...& ...&  ...\\
10 &0.5$^{+0.4}_{-0.3}$& 1.3$^{+0.1}_{-0.1}$&  49.1$^{+5.2}_{-5.4}$&  1.7$^{+1.2}_{-0.2}$& 0.4$^{+0.6}_{-0.3}$&  4.0$^{+4.8}_{-2.0}$\\
11 &1.5$^{+0.6}_{-0.1}$& 0.9$^{+0.1}_{-0.1}$&  11.4$^{+1.3}_{-1.5}$&  1.3$^{+0.1}_{-0.1}$& 0.6$^{+0.1}_{-0.2}$&  5.3$^{+0.5}_{-0.4}$\\
\hline
\end{tabular}
\end{table}
\normalsize


\noindent
{\bf 2MASS\,J12005517} (ID\#10) We find that the spectrum of this object is best reproduced using an extraordinarily high mass fraction of crystalline silicates of $\sim$49\%. This may be a contrast effect: because of the very low luminosity of ID\#10, its 10\mum\ silicate feature arises mainly in the inner $\sim$0.5\,AU of the disk. The central regions of disks can be highly crystalline \citep{2004Natur.432..479V}, possibly leading to very high apparent crystallinities if only the very central disk regions contribute to the part of the spectrum used in the mineralogical analysis \citep[see also ][]{2005Sci...310..834A}. In \fig~\ref{Fig:dust_growth}, we compare the IRS spectrum of ID\#10 in the 7.8--14.3\mum\ spectral range with the  mass absorption coefficients of enstatite and amorphous silicates with different grain sizes. The spectral signature of enstatite is clearly present in the spectrum of ID\#10.

\noindent
{\bf 2MASS\,J12014343} (ID\#11) As discussed in Sect.~\ref{Sec:edg_on}, source ID\#11 may be harboring a disk with a high inclination. Therefore, the IRS spectrum of ID\#11 can be moderately reddened by the cold outer disk. In order to deredden the spectrum, we would need to know the magnitude of the extinction and the proper extinction law at the mid-infrared wavelengths, both of which are not well known. We therefore applied no reddening correction, took the observed spectrum at face value, and fed it to the TLTD routines. Since the TLTD routines only work properly for low-extinction sources, our results for source ID\#11 should be regarded as tentative.

\noindent
{\bf USNO-B120144.7} (ID\#8) The luminosity and spectral type of this source are similar to those of ID\#10. Both sources also show similar accretion activity (see Table~\ref{Tab:acc}) and  similar SEDs (see \fig~\ref{Fig:SED2}). However, in contrast to ID\#10, the IRS spectrum of ID\#8 does not show a silicate feature around 10\mum\ (see \fig~\ref{Fig:dust_growth}). This suggests that the silicates in the warm disk atmosphere of ID\#8 have grown to sizes larger than those in the disk of ID\#10.  In the IRS spectrum of ID\#8, we clearly detect the C$_{2}$H$_{2}$ rovibrational band at $\sim$13.7\mum, but do not find the HCN rovibrational band at $\sim$14.0\mum\ (see \fig~\ref{Fig:dust_growth}). This is consistent with what \citet{2009ApJ...696..143P} discovered in Spitzer spectra of other cool stars, namely, that the C$_{2}$H$_{2}$/HCN flux ratios from cool star disks are large with a median of $\sim$3.8, which is an order of magnitude higher than the median C$_{2}$H$_{2}$/HCN flux ratio in spectra of disks surrounding sun-like stars ($\sim$0.34).

\begin{figure*}[ht]
\centering
\includegraphics[width=1.95\columnwidth]{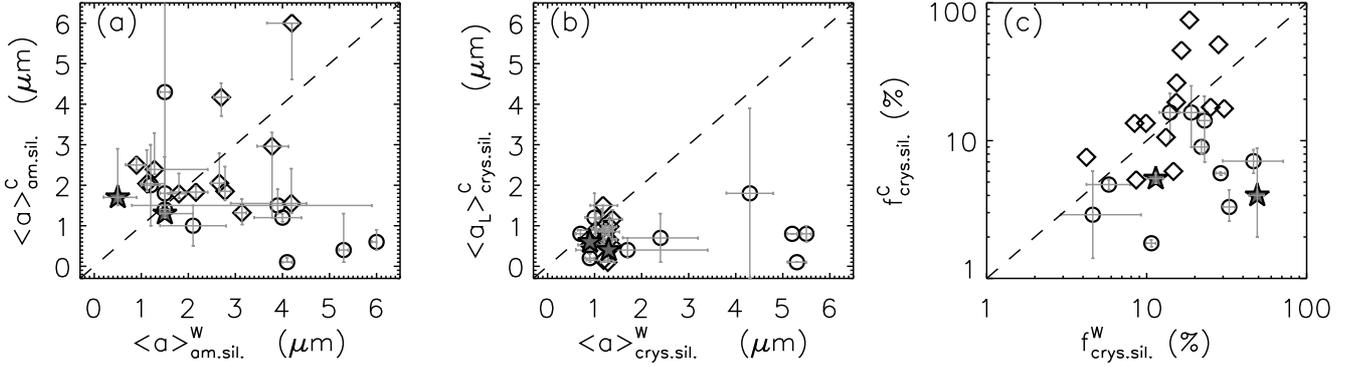}
\caption{(a) Comparison of the mass-averaged sizes of amorphous dust grains as derived from the shorter and longer wavelength part of the IRS spectra (indicated with ``W'' and ``C'' superscripts for ``warm'' and ``cold'' disk regions) of sources in $\epsilon$\,Cha, MBM\,12, $\eta$\,Cha, and the Coronet cluster, and the targets from \citet{2010A&A...520A..39O}. The circles are for the targets in MBM\,12, $\eta$\,Cha, and  the Coronet cluster. The diamonds show the targets from \citet{2010A&A...520A..39O}. The pentagram marks object ID\#10 and 11 in $\epsilon$\,Cha. (b) Similar to (a), but for mass-averaged sizes of crystalline dust grains. (c) Comparison of the mass fraction of crystalline dust grains as derived from the shorter  and longer wavelength part of the IRS spectra. The symbols are similar to panel (a).}\label{Fig:all_dust}
\end{figure*}

\vspace{0.3cm}
\noindent
\noindent \textbf{(b) Dust properties of disks around cool T~Tauri stars} \\
\vspace{-0.3cm}

\noindent
We now combine the dust properties derived for the $\epsilon$\,Cha members with those derived for several other regions and do a statistical analysis of the combined data set. We collected the available literature data for protoplanetary disks surrounding young stars of spectral type M0 or later in the MBM\,12 \citep{2009A&A...497..379M}, $\eta$\,Cha \citep{2009ApJ...701.1188S} associations, the Coronet cluster \citep{2008ApJ...687.1145S}, and the sample in the c2d legacy program \citep{2009A&A...507..327O,2010A&A...520A..39O}. The dust properties for the targets in  MBM\,12,  $\eta$\,Cha, and the Coronet cluster are derived from Spitzer IRS spectra using the same TLTD spectral decomposition routines that we used for the $\epsilon$\,Cha data. As described in Sect.~\ref{Sec:dust_for_epsilon_cha}, the IRS spectra are divided into two regions, a short (7--17\mum) wavelength part tracing the warm inner disk regions and a long (17--37\mum) wavelength part that is more sensitive to cooler regions further from the central star.  For targets in the c2d legacy program,  \citet{2010A&A...520A..39O} performed the spectral decomposition using the B2C method, which includes  continuum emission and two main components, i.e., a warm and a cold one responsible for the 10\mum\ and 20-30\mum\ emission features, respectively. This more or less corresponds to the short  and long  wavelength regime of the IRS spectra in the TLTD spectral decomposition approach.

In \fig~\ref{Fig:all_dust}, we compare the dust properties in the warmer and cooler regions. We compare the average sizes of dust grains for both the amorphous and crystalline grain population, along with the mass fractions of crystalline silicates. As shown in \fig~\ref{Fig:all_dust}(a)(b), the average sizes of the amorphous and crystalline silicates in the warmer regions of protoplanetary disks are substantially larger than those in the cooler disk. This may indicate that dust growth has been more efficient closer to the central star than in regions at larger distances. However it may also mean that the disks are more turbulent in the warm inner regions than further out, which allows larger grains to remain visible in the disk surface instead of settling to the midplane. A combination of both effects is also plausible. Figure~\ref{Fig:all_dust}(c) shows that the dust in the warm inner regions of protoplanetary disks generally contains a higher fraction of crystalline material than the dust in cooler regions. This is consistent with earlier findings \citep[e.g.,][]{2004Natur.432..479V,2009A&A...497..379M}.

\begin{figure}[ht]
\centering
\includegraphics[width=0.95\columnwidth]{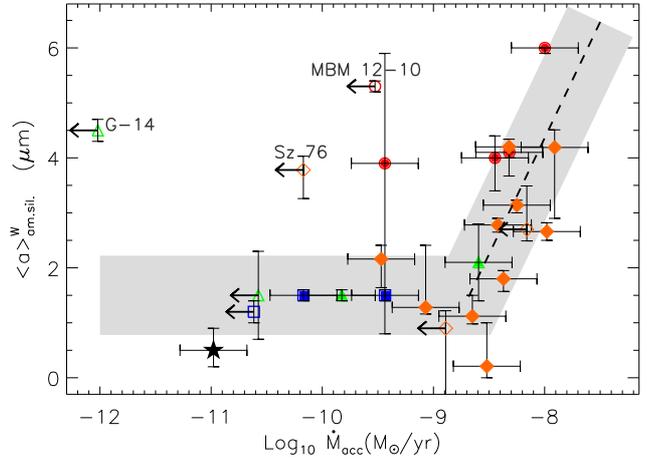}
\caption{The average  sizes of amorphous grains plotted versus the accretion rate of the central stars. The grain sizes are derived in the warmer region of the IRS spectra. The filled symbols show the CTTSs, and the open symbols represent the WTTSs. {\newnewrev The targets in MBM\,12  are shown as  circles, in $\eta$\,Cha as squares, and in the Coronet cluster as  triangles.} The diamonds show the targets from \citet{2010A&A...520A..39O}. The pentagram marks object ID\#10 in $\epsilon$\,Cha. The thick dashed lines are the fit to the relation between the accretion rates and the grain sizes for $\dot{M}_{acc}>$10$^{-9}$\,\accunit.}\label{Fig:size_acc}
\end{figure}

We estimated the accretion rates for all stars from their H$\alpha$ line luminosity using the method described in Sect.~\ref{Sec:Accretion} and list these in Table~\ref{Tab:dust_for_all}. In \fig~\ref{Fig:size_acc}, we plot the average grain sizes of amorphous silicates against accretion rates.  The data suggest a positive correlation between the average grain sizes of the amorphous silicates and the accretion rates if the latter is above $\sim$10$^{-9}$\,\accunit\ (see \fig~\ref{Fig:size_acc}). Below this value, the grains in the majority of disks are small, independently of the accretion rate.  In order to see whether there is any correlation between both observables, we apply a Kendall $\tau$ test. If two datasets are fully correlated, this test returns a value of $\tau$\,$=$\,1. If they are anti-correlated, we get $\tau$\,$=$\,$-$1, and if they are independent, we obtain $\tau$\,$=$\,0. The Kendall $\tau$ test also returns a probability $p$, which is smaller when the correlation is more significant. We use the Kendall $\tau$ test to evaluate the possible correlation between the average grain sizes of amorphous silicates and the accretion rates, which yields $\tau$=0.79 and $p$=0.006 for  $\dot{M}_{acc}>10^{-9}$\,\accunit, and  $\tau$=0.16 and $p$=0.58 for  $\dot{M}_{acc}\le10^{-9}$\,\accunit. Thus there is a significant correlation between  the average grain sizes of amorphous silicates and the accretion rates when  $\dot{M}_{acc}>10^{-9}$\,\accunit. A possible explanation for this relation is that both accretion and the presence of large grains in the disk surface require some level of turbulence in the disk. Thus, both a large average grain size and a high accretion rate are tale tell signs of a turbulent disk, though there need not be a direct causal connection between both observables.

There are three outliers in \fig~\ref{Fig:size_acc}, MBM\,12-10, G-14, and Sz\,76, which  are all classified as WTTSs due to their small H$\alpha$ EWs. Yet, all of them show large average grain sizes in their IRS spectra. {\newnewrev  G-14 shows a quite globally depleted SED, which can be due to strong overall dust settling and/or grain growth in the disk \citep{2008ApJ...687.1145S}. MBM12-10 and Sz 76 can be classified as transition disks \citep{2009A&A...497..379M,2010ApJ...724..835W}. It is still unknown which mechanisms are responsible for the evolution from normal disks to transition disks. One of the proposed mechanisms is dust growth and settling, which is believed to work in some of the transition disks in the Cep OB2 region  \citep{2011ApJ...742...39S}. However, in the Chamaeleon I star-forming region, \citet{2011ApJS..193...11M} found that some of the transition disks show 10\mum\ features from relatively unprocessed grains compared with normal disks. This indicates that, at least in disk surfaces, the dust population has not undergone substantial grain growth in these objects. Thus that a transition disk geometry does not necessarily go hand in hand with large silicate grains dominating the mid-infrared spectra. Nevertheless, the deviant behavior of sources MBM12-10, G-14, and Sz\,76 from the correlation between the average grain size of the amorphous grains and the accretion rate is plausibly due to their special disk structures (with inner holes) or extreme grain growth  \citep[see also][]{2011ApJ...742...39S}.}

\citet{2007ApJ...659.1637S} have discovered a correlation between the average grain sizes of silicates and accretion rates that is similar to what we find for accretion rates above 10$^{-9}$\,\accunit. In addition, they argue that turbulence in disks prevents large grains from settling into the disk interior, where they are invisible to Spitzer, and also promotes accretion, thereby leading to a correlation between both observables. What is new in the current study is that the correlation between the accretion rates and grain sizes  breaks down for $\dot{M}_{acc}<$10$^{-9}$\,\accunit, suggesting that the turbulence required to sustain such accretion rates is insufficient to support large grains against settling. We fit the relation between the average grain size of the amorphous dust and the accretion rate with the following curve:

\[\langle a \rangle_{\rm am.sil.}^{\rm W} = \left\{ 
\begin{array}{l l}
  38.2+4.2\times Log \dot{M}_{acc}& \quad \mbox{if $\dot{M}_{acc}>10^{-9}$\,\accunit}\\
  \sim1.5& \quad \mbox{if $\dot{M}_{acc}\le10^{-9}$\,\accunit.}\\ \end{array} \right. \]

\noindent
Here, $\langle a \rangle_{\rm am.sil.}^{\rm W}$  (in\mum) is the average size of amorphous silicates in the warmer disk regions, and \Mdotacc\ is the accretion rate in\,\accunit. 

{\newnewrev In \fig~\ref{Fig:size_acc}, disks in younger MBM\,12 show more processed dust grains than those in relatively older $\eta$\,Cha. Since younger disks tend to show more processed dust grains, we cannot explain the positive correlation between $\langle a \rangle_{\rm am.sil.}^{\rm W}$ as  a global effect of disk evolution. We would expect older disks to show bigger dust grains because dust grains are expected to grow with disk evolution.}

\begin{figure}
\centering
\includegraphics[width=0.95\columnwidth]{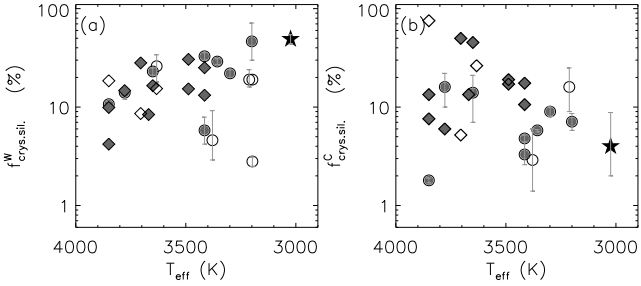}
\caption{The mass fractions of crystalline material compared to the effective temperature of the central stars (which correlates closely with the bolometric luminosity, see \fig\ref{Fig:HRD}). The crystallinities are derived from the 7--17\mum\ part of the IRS spectra.  (b) similar to (a), except that the crystallinities are derived from in 17--37\mum\ part of the IRS spectra. The symbols are identical to those in \fig~\ref{Fig:size_acc}. }\label{Fig:fcrys_teff}
\end{figure}

Figure~\ref{Fig:fcrys_teff} shows the mass fraction of crystalline material in the ``warmer'' and ``colder'' disk regions versus the effective temperatures of the central stars for our sample of cool T-Tauri stars. We use the Kendall $\tau$ test to evaluate any possible correlation between both quantities, which yields $\tau$=$-$0.25 and $p$=0.07 for the warm disk part (panel a in \fig~\ref{Fig:fcrys_teff}), and $\tau$=0.19 and $p$=0.19 for the cooler disk part (panel b  in \fig~\ref{Fig:fcrys_teff}). Thus there is no significant correlation between the mass fraction of crystalline silicates in the disk and the stellar effective temperature for cool T~Tauri stars.

\begin{figure}
\centering
\includegraphics[width=0.95\columnwidth]{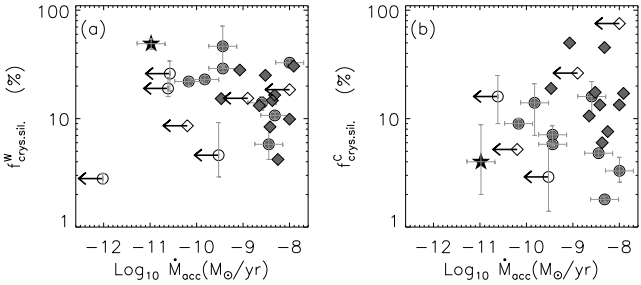}
\caption{The mass fractions of crystalline silicates plotted versus the accretion rates of the central objects. Panel (a) shows the crystallinities as derived from the 7--17\mum\ part of the IRS spectra, panel (b) shows the same quantity derived in the 17--37\mum\ spectral region. The symbols are identical to those in \fig~\ref{Fig:size_acc}.}\label{Fig:fcrys_acc}
\end{figure}

In \fig~\ref{Fig:fcrys_acc}, we plot the crystallinity of the disk material against the mass accretion rates. Once again, we run the Kendall $\tau$ test, and find essentially no significant correlation in the warm disk regions ($\tau$=$-$0.27, $p$=0.10) and the colder parts of the disks ($\tau$=$-$0.02, $p$=0.88). {\rev \citet{2009A&A...508..247G} suggested that irradiation of dust grains by energetic ions from the stellar winds of young stars can amorphize the surface layer of the protoplanetary dust very efficiently, thus erasing any correlation between crystalline mass fraction and stellar parameters, such as bolometric luminosity, effective temperature, accretion rate, or disk geometry. \citet{2009Natur.459..224A} discovered that episodes of increased accretion may create new crystals. If strongly variable accretion is characteristic of T~Tauri stars, as suggested by, e.g. \citet{2011MNRAS.411L..51M} and Fang et al. (submitted to ApJs), this may also account for the lack of obvious correlations between any disk or stellar properties and  crystallinity of the dust in the disk surface.}

\section{Summary}\label{Sec:summary}
We have studied the members of the $\epsilon$\,Cha association, focusing on the properties of the central stars and their circumstellar disks. We used a combination of available archive data, our own Spitzer spectroscopy, and VLT/VISIR imaging data.

Using proper motions, we confirm the membership of most studied stars but question the membership of stars CXOU\,J120152.8 (ID\#9) and 2MASS\,J12074597(ID\#12). We estimated the masses and ages of the $\epsilon$\,Cha members and find HD\,104237C to be a sub-stellar object with a very low mass of 13--15\,\Mjup, putting it at the boundary between brown dwarfs and ``free-floating planets''.

The object 2MASS\,J12014343 (ID\#11) is unusually faint at optical wavelengths and also shows exceptionally large EWs of some optical emission lines and an exceptionally strong infrared excess. Similar objects have been discovered in other star-forming regions. We tentatively explain this behavior with a flared disk seen at moderately high inclination in which the cold outer disk regions cause sufficient extinction to effectively screen the central star, which is then seen mostly in scattered light, but allow most of the infrared light from the warm disk regions to pass. The protoplanetary disks surrounding the cool stars USNO-B120144.7 and 2MASS\,J12005517 show evidence of a reduced height of the optically thick disk due to dust settling. The disk around HD\,104237E shows evidence of partial dissipation in its inner part, while its outer disk remains essentially intact. 

We found that both disk frequency and accretor frequency in the $\epsilon$\,Cha association are higher than those in relatively dense clusters of similar age. Five other sparse stellar associations for which data are available in the literature also show comparatively high disk frequencies. Disk evolutions appears to proceed substantially more slowly in sparse associations compared to denser environments. In addition, the disk frequencies in  sparse associations are almost constant at ages less than $\sim$4\,Myr.

{\rev The 13.7\mum\ rovibrational band of C$_{2}$H$_{2}$  is detected in the IRS spectrum of USNO-B120144.7.}  

We derive the mineralogical composition and grain size distribution of the (sub-) micron-sized dust in the disk atmosphere using the TLTD method. We find that the average grain sizes and fractions of crystalline material are higher in the warm inner disk regions that dominate the short wavelength part of the IRS spectra compared to cooler regions at a longer distance from the central star that contribute mostly to the longer wavelength range covered by the IRS. We also find that the average sizes of amorphous grains in the warm inner disk regions show a positive correlation with the accretion rates  if the latter is higher than $\sim$10$^{-9}$\,\accunit.

\begin{acknowledgements} 
 Many thanks to V.~Roccatagliata for the useful discussions on this paper and to the anonymous referee for comments that help to improve this paper. MF acknowledges the support by NSFC through grants 11203081, 10733030 and 11173060. ASA acknowledges support from the "Ramon y Cajal" Program from the Spanish MICINN/MINECO.  This research has made use of the SIMBAD database, operated at CDS, Strasbourg, France. This publication makes use of data products from the Two Micron All Sky Survey, which is a joint project of the University of Massachusetts and the Infrared Processing and Analysis Center/California Institute of Technology, funded by the National Aeronautics and Space Administration and the National Science Foundation. This publication makes use of data products from the Wide-field Infrared Survey Explorer, which is a joint project of the University of California, Los Angeles, and the Jet Propulsion Laboratory/California Institute of Technology, funded by the National Aeronautics and Space Administration. This research is based on observations with AKARI, a JAXA project with the participation of ESA. This work is in part  based  on observations made with the Spitzer Space Telescope, which is operated by the Jet Propulsion Laboratory, California Institute of Technology under a contract with NASA. 
\end{acknowledgements} 

\begin{appendix}
\section{The disk frequencies in star-formation regions}\label{Appen:disk_fractions}
In Table~\ref{Tab:disk_frequency}, we list each star-formation region (SFR) used in \fig~\ref{Fig:df}. In total, 16 SFRs are included with median ages ranging from $\sim$1.5 to $\sim$13\,Myr.  Most of the SFRs in the tables have an age estimate using the PMS evolutionary tracks from \citet{1998A&A...337..403B} in the literature. The references for these ages are also listed in the table. The 5 SFRs, i.e., $\epsilon$~Cha, TW~Hya, L988e, Tr~37, and NGC~7160, are without ages estimated from the tracks of \citet{1998A&A...337..403B}. We estimate their ages using these tracks with a  mixing length parameter $l/H_{\rm P}$=1. In Table~\ref{Tab:disk_frequency}, we give the disk frequency for each SFR as well as the references. We corrected these disk frequencies by removing the transitional disk objects that  only show excess emission at wavelengths longer than 8\mum. For the disk frequencies estimated in this work, we distinguish the disk population from diskless ones using the SED slope criteria from \citet{2006AJ....131.1574L}. The disk frequencies in the sparse stellar association, MBM\,12, $\epsilon$\,Cha, $\eta$\,Cha, and TW\,Hya tend to be systematically above those of the other SFRs with similar ages.

\begin{figure}
\centering
\includegraphics[width=0.95\columnwidth]{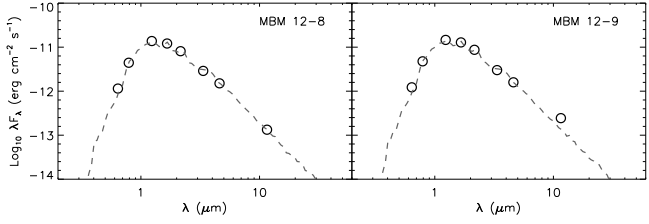}
\caption{SEDs of  two stars in MBM\,12 associations. The photospheric emission level is indicated with a grey dashed curve in each panel. The open circles show the photometry in different bands.}\label{Fig:SED_MBM12}
\end{figure}

{\bf MBM\,12}~~  In this association, there are 12 known members with K to M spectral types. Among these, eight objects (MBM\,12-1,2,3,4,5,6,10,12) were detected by the Spitzer IRS, and seven objects show evidence for a circumstellar disk (MBM\,12-2,3,4,5,6,10,12). Three of the seven objects are suggested to be transition disk candidates\citep{2009A&A...497..379M}. {\rev In the recently published catalog from the Wide-field Infrared Survey Explorer \citep[WISE,][]{2010AJ....140.1868W} the four objects lacking IRS spectra (MBM12-7,8,9,11) are detected in three photometric bands of WISE: W1 (3.4\mum), W2 (4.6\mum), and W3 (12\mum) bands. With the new infrared photometry, we identify only MBM12-9 (see Fig.~\ref{Fig:SED_MBM12}) as having an IR excess at 12\mum, whereas the other three sources have WISE fluxes  consistent with the stellar photospheric emission. The object MBM12-8 shows strong H$\alpha$ emission (EW$_{\rm H\alpha}$=$-$120$\AA$), pointing at active accretion. However it shows no excess emission at 12\mum\ and shorter wavelengths, which indicates that the warm inner disk regions have already dissipated. Possibly this is a transition disk object with a large opacity hole in the inner disk, where some gas is still present. This allows accretion to proceed. Observations at longer wavelengths are needed to confirm the transition disk nature of MBM12-8. For the time being, we do not consider the object to have a confirmed circumstellar disk.}  Thus, the disk frequency for all known members in MBM\,12 is {\newnewrev $\sim$67$^{+10}_{-15}$\%} (8/12), while for members with masses of 0.1--0.6\,\Msun, it is  $\sim$ {\newnewrev 67$^{+11}_{-17}$\%} (6/9).  

\begin{figure}
\centering
\includegraphics[width=0.95\columnwidth]{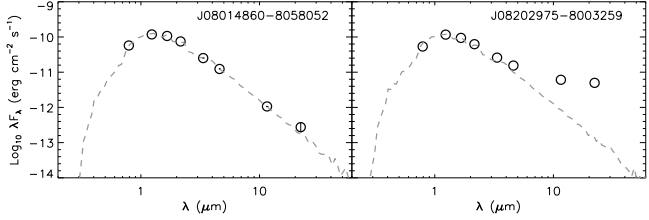}
\caption{SEDs of  two stars in the $\eta$~Cha association. The photospheric emission level is indicated with a grey dashed curve in each panel. The open circles show the photometry in different bands.}\label{Fig:SED_eta_cha}
\end{figure}

{\bf $\eta$\,Cha}~~ In the $\eta$\,Cha association, 18 members have been observed with Spitzer. The disk frequency is estimated to be   {\newnewrev $\sim$44$^{+12}_{-11}$\%} (8/18) \citep{2005ApJ...634L.113M,2009ApJ...701.1188S}. {\newnewrev The sources RECX-3 and RECX-4 only show excess emission at wavelengths longer than 10\mum. In this work, we restrict our disk census to the Spitzer IRAC bands, thus excluding the two sources when calculating the disk fraction in  $\eta$\,Cha.} \citet{2010MNRAS.406L..50M} proposed that seven candidate low-mass members of $\eta$\,Cha comprise a halo surrounding the cluster core. Two of these may have disks that are actively accreting, as indicated by their H$\alpha$ emission \citep{2010MNRAS.406L..50M}. We have found all seven objects in the WISE catalogue. Only one of these (J08202975-8003259) shows excess emission in the WISE data (see Fig.~\ref{Fig:SED_eta_cha}).  The source 2MASS~J08014860-8058052 does not show clear excess emission at wavelengths up to 22\mum\ (see Fig.~\ref{Fig:SED_eta_cha}).  Thus, we presently consider a possible disk around the latter object to be confirmed. Including the seven new members, the  disk frequency of $\eta$\,Cha is  {\newnewrev $\sim$28$^{+10}_{-7}$\%} (7/25). For the members with masses of 0.1--0.6\,\Msun, the disk frequency is {\newrev $\sim$32$^{+12}_{-8}$}(6/19). 

{\bf TW Hya}~~ To estimate the disk frequency of this association, we first constructed a catalog of members of this group. Our catalog is based on the TW\,Hya membership criteria as refined by \citet{2005ApJ...634.1385M}, and includes TWA-1, 2A, 2B, 3A, 3B, 4, 5A, 5B, 6, 7, 8A, 8B, 9A, 9B, 10, 11A, 11B, 13A, 13B, 14, 15A, 15B, 16, 20, 21, 23, 25, 26, 27, 28, plus new identified members TWA-29 30A, 30B \citep{2007ApJ...669L..97L,2010AJ....140.1486L,2010ApJ...714...45L}. Among these members, TWA-1, 3A, 4, 7, 11A, 27, 28, and 30B show evidence for disks {\newnewrev with excess emission at wavelengths shorter than 8\mum\ \citep{1999ApJ...521L.129J,2008ApJ...681.1584R,2010AJ....140.1486L,2010ApJ...714...45L}. TWA-30A may harbor a  disk because its  optical and infrared spectra show emission lines due to accretion activity \citep{2010AJ....140.1486L}. However, there is no infrared data to suggest whether TWA-30A shows excess emission at wavelengths shorter than  8\mum. Thus, we do not count this source when calculating the disk fraction in TW Hya. The disk frequency for all known members is  estimated to be   {\newnewrev $\sim$25$^{+9}_{-6}$\%}(8/32), and the disk frequency for members with masses of 0.1--0.6\,\Msun\ is    {\newnewrev $\sim$19$^{+13}_{-6}$\%} (3/16)}.   

\renewcommand{\tabcolsep}{0.03cm}
\begin{table}
\caption{\label{Tab:disk_frequency} Fractions of YSO with excess emission at wavelengths less than $\sim$8\mum\ in different SFRs. Column 1: $^{a}$ Sparse stellar associations.  Column 2: the median age of each SFR estimated from \citet{1998A&A...337..403B}.  Column 3: the references for the median age of SFR. Column 4: the disk frequencies for all known members. Column 5: the references for disk frequencies of all known members.  Column 6: the disk frequencies for the members with masses of 0.1-0.6\Msun.  Column 7: the references for disk frequencies listed in Column 6.
}
 \centering
\begin{tabular}{lcccccccccc}
\hline\hline
(1)                    &(2)    &(3)        &(4)             &(5)   &(6)                & (7)\\
                       &Age    &           &DF$_{\rm All}$   &      &DF$_{0.1-0.6M_{\odot}}$ &    \\
SFR                    &(Myr)  &Ref        &(\%)            & Ref  &(\%)               & Ref\\
\hline     

 MBM\,12$^{a}$          &  2.0   &(1)       &67$^{+10}_{-15}$ &(2)    &67$^{+17}_{-11}$  &(2,3)\\ 
$\epsilon$\,Cha$^{a}$   &  4.0    &(3)       &55$^{+13}_{-15}$ &(3)    &50$^{+16}_{-16}$  &(3)\\ 
$\eta$\,Cha$^{a}$       &  6.0   &(1)       &28$^{+10}_{-7}$  &(4,5,6,7) &32$^{+12}_{-8}$ &(4,5,6,7,3)\\
TW\,Hya$^{a}$           &  9.0   &(8, 3)    &25$^{+9}_{-6}$    &(3)    &19$^{+13}_{-6}$    &(3)\\
CrA$^{a}$               &  1.5   &(9)      &70$^{+7}_{-9}$       &(10)    &\nodata          &\nodata\\ 
Taurus$^{a}$            &  1.5   &(11)      &62$\pm$4        &(12)     &58$\pm$6         &(12,3)\\
L988e                   &  1.5   &(13, 3)   &79$\pm$8         &(14)     &\nodata          &\nodata\\
NGC\,2068/71            &  1.5   &(15)      &71$\pm$7         &(15)     &63$\pm$8         &(15,3)\\
Cha\,I                  &  2.0   &(16)      &53$\pm$5         &(17)     &44$\pm$6         &(17,3)\\ 
IC\,348                 &  2.5  &(18)       &41$\pm$4         &(19)     &40$\pm$5         &(19,3)\\
NGC\,2264               &  2.5  &(20)       &38$\pm$3         &(21)    &\nodata          &\nodata\\
$\sigma$\,Ori           &  2.5  &(22)       &37$\pm$3         &(23)     &33$\pm$4           &(23, 3)\\ 
 Tr\,37                 &  3.0  &(24, 3)    &39$\pm$6        &(25, 26)    &\nodata       &\nodata\\
$\lambda$\,Ori          &  5.0  &(27)       &15$\pm$3        &(28, 29)    &18$\pm$5        &(28, 29, 3) \\
NGC\,7160               &  13.0   &(25, 3)  &4$\pm$3        &(26)    &\nodata          &\nodata\\
NGC\,2362               &  5.0  &(30)       &13$\pm$2        &(31, 3)    &21$\pm$4          &(31,3)\\
Ori\,OB1b               &  6.0  &(32, 33)   &16$\pm$4        &(34, 3)    &15$\pm$4         &(34,3) \\
25~Ori                  &  7.4  &(32, 33)   &5$\pm$3         &(34, 3)    &5$\pm$2          &(34,3) \\
Upper\,Sco              &  5.0   &(35)      &13$\pm$2         &(36)     &20$\pm$5         &(36,3) \\
\hline
\end{tabular}
\tablebib{(1) \citet{2004ApJ...609..917L}; (2) \citet{2009A&A...497..379M}; (3) this paper; (4) \citet{2005ApJ...634L.113M}; (5) \citet{2009ApJ...701.1188S}; (6) \citet{2010MNRAS.406L..50M}; (7) \citet{2011MNRAS.411L..51M}; (8) \citet{1999ApJ...512L..63W}; (9) \citet{2011ApJ...736..137S}; (10) \citet{2008ApJ...687.1145S}; (11) \citet{2002ApJ...580..317B}; (12) \citet{2010ApJS..186..111L}; (13) \citet{2006AJ....131.1530H}; (14) \citet{2008ApJ...675..491A}; (15)  \citet{2009A&A...504..461F}; (16) \citet{2007ApJS..173..104L}; (17) \citet{2008ApJ...675.1375L}; (18) \citet{2007AJ....134..411M}; (19)  \citet{2006AJ....131.1574L}; (20)  \citet{2006A&A...455..903F}; (21) \citet{2009AJ....138.1116S}; (22)  \citet{2004AJ....128.2316S}; (23) \citet{2008ApJ...688..362L}; (24) \citet{2005AJ....130..188S}; (25) \citet{2006ApJ...638..897S}; (26) \citet{2009AJ....138....7M}; (27) \citet{2007ApJ...664..481B}; (28) \citet{2010ApJ...722.1226H}; (29) \citet{2007ApJ...664..481B}; (30)  \citet{2001ApJ...563L..73M}; (31) \citet{2009ApJ...698....1C}; (32) \citet{2005AJ....129..907B}; (33) \citet{2007ApJ...661.1119B}; (34) \citet{2007ApJ...671.1784H}; (35) \citet{2002AJ....124..404P}; (36) \citet{2006ApJ...651L..49C}.}
\end{table}
\normalsize

\section{The dust properties around M-type PMS stars in other regions}\label{Appen:dust}

In Table~\ref{Tab:dust_for_all}, we list the targets used in Figs.~\ref{Fig:all_dust}, ~\ref{Fig:size_acc}, ~\ref{Fig:fcrys_teff}, and ~\ref{Fig:fcrys_acc}. The total luminosity, extinction, and accretion rate listed here for each object are derived using the procedure described in Sect.~\ref{Sec:targets} and in \citet{2009A&A...504..461F}. The stellar masses and ages are estimated  by comparison to  theoretical PMS evolutionary tracks from \citet{2008ApJS..178...89D}. We collected the dust properties for each disk from the literature \citep{2008ApJ...687.1145S,2009A&A...497..379M,2009ApJ...701.1188S,2010A&A...520A..39O}. These include the average sizes of amorphous and crystalline silicate grains, along with the mass fractions of crystalline silicate grains, and were derived using either the same TLTD method that we used for the analysis of the spectra of $\epsilon$\,Cha members (as described in Sect.~\ref{Sec:IRS}) or the B2C method, which is similar to the TLTD method. 

\renewcommand{\tabcolsep}{0.03cm}
\begin{table*}
\caption{Stellar and disk properties for young stellar objects. Column 10: the accretion rates are derived from the H$\alpha$ emission line (see Sect.~\ref{Sec:Accretion}), besides objects SX\,Cha, WX\,Cha, and XX\,Cha, whose accretion rates are from \citet{1998ApJ...495..385H}. Columns 13, 14, 15: the average sizes of amorphous ($\langle$a$\rangle$$_{\rm am.sil.}^{\rm W}$) and crystalline grains ($\langle$a$\rangle$$_{\rm cryst.sil.}^{\rm W}$) and the mass fractions of crystalline grains (f$_{\rm cryst}^{\rm W}$) are derived from fitting silicate features around 10\mum\ with the TLTD or B2C methods. Columns 16, 17, 18: the average sizes of amorphous ($\langle$a$\rangle$$_{\rm am.sil.}^{\rm C}$) and crystalline grains ($\langle$a$\rangle$$_{\rm cryst.sil.}^{\rm C}$) and the mass fractions of crystalline grains (f$_{\rm cryst}^{\rm C}$) are derived from fitting silicate features around 20--30\mum\ with the TLTD or B2C methods. \label{Tab:dust_for_all}} 
\scriptsize
\centering
\begin{tabular}{lccccccccccccccccccc}
\hline\hline
(1) &(2)&(3)&(4)&(5)&(6)&(7)&(8)&(9)&(10) &(11)&(12)&(13)&(14)&(15)&(16)&(17)&(18)&(19)\\
    &RA          & DEC           &      &Teff  &L$_{\rm bol}$ &Av    &H$\alpha$& & Log~$\dot{\rm M}$$_{acc}$ &Mass &Age &$\langle$a$\rangle$$_{\rm am.sil.}^{\rm W}$&$\langle$a$\rangle$$_{\rm cryst.sil.}^{\rm W}$ &f$_{\rm cryst}^{\rm W}$  &$\langle$a$\rangle$$_{\rm am.sil.}^{\rm C}$&$\langle$a$\rangle$$_{\rm cryst.sil.}^{\rm C}$ &f$_{\rm cryst}^{\rm C}$ &\\
Object        &(J2000)     & (J2000)       & Spt  &(K)   &(L$\odot$)   &(mag)&($\AA$)&class &(M\accunit)&(\Msun) &(Myr)&(\mum)&(\mum) &(\%) &(\mum)&(\mum) &(\%) &Ref\\ 
\hline
  \multicolumn{19}{c}{TLTD method}\\
\hline
MBM\,12-2 &02:56:07.99 &+20:03:24.3 &M0 &3850  &0.476  &0.5 &-40.0  &C  &-8.32 &0.57 &1.8  &4.1$^{+0.1}_{-0.1}$  &5.2$^{+0.1}_{-0.1}$  &10.7$^{+0.7}_{-0.8}$  &0.1$^{+0.1}_{-0.0}$  &0.8$^{+0.0}_{-0.0}$  &1.8$^{+0.1}_{-0.1}$   &(1)\\
MBM\,12-3 &02:56:08.42 &+20:03:38.6 &M3 &3415  &0.636  &0.0 &-25.0  &C  &-8.45 &0.27 &0.2  &4.0$^{+0.4}_{-0.6}$  &1.7$^{+1.7}_{-1.1}$  &5.8$^{+2.1}_{-1.6}$  &1.2$^{+0.1}_{-0.1}$  &0.4$^{+0.3}_{-0.2}$  &4.8$^{+0.4}_{-0.4}$   &(1)\\
MBM\,12-6 &02:58:16.09 &+19:47:19.6 &M5 &3200  &0.204  &0.0 &-29.0  &C  &-9.44 &0.21 &0.6  &3.9$^{+2.0}_{-3.1}$  &5.3$^{+0.2}_{-0.2}$  &46.5$^{+24.9}_{-16.6}$  &1.5$^{+0.4}_{-0.2}$  &0.1$^{+0.0}_{-0.0}$  &7.1$^{+1.5}_{-1.3}$   &(1)\\
MBM\,12-10 &02:58:21.10 &+20:32:52.7 &M3.25 &3379  &0.280  &0.0 &-12.0  &W  &$<$-9.53 &0.28 &0.8  &5.3$^{+0.1}_{-0.1}$  &1.9$^{+2.1}_{-0.9}$  &4.6$^{+4.6}_{-1.7}$  &0.4$^{+0.9}_{-0.3}$  &\nodata$^{}_{}$  &2.9$^{+3.1}_{-1.5}$   &(1)\\
MBM\,12-12 &03:02:21.05 &+17:10:34.2 &M3 &3415  &0.563  &0.0 &-69.0  &C  &-8.00 &0.27 &0.3  &6.0$^{+0.0}_{-0.1}$  &5.5$^{+0.1}_{-0.1}$  &32.8$^{+2.9}_{-2.8}$  &0.6$^{+0.3}_{-0.2}$  &0.8$^{+0.2}_{-0.2}$  &3.3$^{+1.1}_{-0.7}$   &(1)\\
CRA-466 &19:01:18.93 &$-$36:58:28.2 &M2 &3650  &0.154  &6.4 &-14.5  &C  &-9.82 &0.54 &7.8  &1.5$^{+0.1}_{-0.1}$  &4.3$^{+0.5}_{-0.5}$  &23.0$^{+2.0}_{-2.0}$  &4.3$^{+2.9}_{-2.9}$  &1.8$^{+2.1}_{-2.1}$  &14.0$^{+7.0}_{-7.0}$   &(2)\\
CRA-4107 &19:02:54.65 &$-$36:46:19.1 &M4.5 &3198  &0.029  &0.0 &\nodata  &...  &\nodata &0.19 &10.9  &5.7$^{+2.6}_{-2.6}$  &4.1$^{+3.3}_{-3.3}$  &19.0$^{+11.0}_{-11.0}$  &\nodata$^{}_{}$  &\nodata$^{}_{}$  &\nodata$^{}_{}$   &(2)\\
G-14 &19:02:12.02 &$-$37:03:09.3 &M4.5 &3198  &0.016  &1.1 &-7.4  &W  &$<$-12.02 &0.18 &22.9  &4.5$^{+0.2}_{-0.2}$  &0.2$^{+0.4}_{-0.4}$  &2.8$^{+0.3}_{-0.3}$  &\nodata$^{}_{}$  &\nodata$^{}_{}$  &\nodata$^{}_{}$   &(2)\\
G-85 &19:01:33.86 &$-$36:57:44.8 &M0 &3778  &0.400  &15.5 &-31.0  &C  &-8.59 &0.54 &2.0  &2.1$^{+0.7}_{-0.7}$  &2.4$^{+0.8}_{-0.8}$  &14.0$^{+2.0}_{-2.0}$  &1.0$^{+0.5}_{-0.5}$  &0.7$^{+0.6}_{-0.6}$  &16.0$^{+6.0}_{-6.0}$   &(2, 3)\\
G-87 &19:01:32.33 &$-$36:58:03.0 &M1.5 &3633  &0.145  &13.7 &-4.0  &W  &$<$-10.58 &0.53 &8.2  &1.5$^{+0.8}_{-0.8}$  &0.7$^{+1.2}_{-1.2}$  &26.0$^{+8.0}_{-8.0}$  &\nodata$^{}_{}$  &\nodata$^{}_{}$  &\nodata$^{}_{}$   &(2, 3)\\
J0843 &08:43:18.58 &$-$79:05:18.2 &M3.4 &3357  &0.080  &0.0 &-90.0  &C  &-9.44 &0.30 &6.2  &1.5$^{+0.1}_{-0.1}$  &0.7$^{+0.1}_{-0.1}$  &29.0$^{+2.0}_{-2.0}$  &1.4$^{+0.1}_{-0.1}$  &0.8$^{+0.1}_{-0.1}$  &5.8$^{+0.1}_{-0.2}$   &(4)\\
RECX-5 &08:42:27.11 &$-$78:57:47.9 &M3.8 &3299  &0.061  &0.0 &-35.0  &C  &-10.17 &0.26 &6.8  &1.5$^{+0.1}_{-0.1}$  &0.9$^{+0.1}_{-0.1}$  &22.0$^{+2.0}_{-2.0}$  &1.8$^{+0.9}_{-0.3}$  &0.2$^{+0.1}_{-0.1}$  &9.0$^{+1.0}_{-1.0}$   &(4)\\
RECX-9 &08:44:16.38 &$-$78:59:08.1 &M4.4 &3212  &0.092  &0.0 &-10.0  &W  &$<$-10.62 &0.21 &2.4  &1.2$^{+0.2}_{-0.2}$  &1.0$^{+0.1}_{-0.2}$  &19.0$^{+5.0}_{-3.0}$  &2.0$^{+1.0}_{-1.0}$  &1.2$^{+0.6}_{-0.4}$  &16.0$^{+9.0}_{-7.0}$   &(4)\\
\hline
  \multicolumn{19}{c}{B2C method}\\
\hline
Sz~50           &13:00:55.32 &$-$77:10:22.2 &M3   &3415  &0.527&1.7 &-29   &C&-8.52 & 0.28 &0.3 &$0.2_{-0.2}^{+0.8}$ &$1.2_{-0.3}^{+0.3}$ &25.1 &\nodata            &\nodata  &17.5 &(5), (6)\\ 
Sz~52           &13:04:24.90 &$-$77:52:30.3 &M2.5 &3488  &0.117&2.9 &-46   &C&-9.47 & 0.40 &6.3 &$2.2_{-0.5}^{+0.3}$ &$1.3_{-0.2}^{+0.2}$ &15.3 &$1.8_{-0.1}^{+0.1}$  &$0.1_{-0.0}^{+0.7}$  &19.0 &(5), (6)\\ 
Sz~62           &13:09:50.37 &$-$77:57:24.0 &M2.5 &3488  &0.363&1.1 &-150  &C&-7.91 & 0.33 &0.8 &$4.2_{-1.3}^{+0.3}$ &$1.3_{-0.1}^{+0.2}$ &30.5 &$1.6_{-0.2}^{+0.9}$  &$1.0_{-0.3}^{+0.3}$  &17.1 &(5), (6)\\ 
IRAS~12535-7623 &12:57:11.78 &$-$76:40:11.5 &M0   &3850  &0.951&2.4 &-15   &C&-8.25 & 0.49 &0.5 &$3.1_{-0.1}^{+0.1}$ &$1.0_{-0.4}^{+0.4}$ &4.2  &$1.3_{-0.3}^{+0.3}$  &$0.9_{-0.3}^{+0.4}$  &7.6  &(5), (6)\\ 
SX~Cha          &10:55:59.76 &$-$77:24:40.1 &M0.5 &3778  &0.294&0.6 &-26.7 &C&-8.37 & 0.58 &3.6 &$1.8_{ -0.2}^{+0.2}$&$1.2_{-0.3}^{+0.3}$ &14.7 &$1.8_{-0.1}^{+0.5}$  &$1.5_{-0.2}^{+0.0}$  &6.0  &(7), (6)\\ 
WX~Cha          &11:09:58.74 &$-$77:37:08.9 &M1.25&3669  &0.878&2.0 &-65.5 &C&-8.42 & 0.38 &0.3 &$2.8_{-0.1}^{+0.1}$ &$1.2_{-0.3}^{+0.3}$ &8.4  &$1.9_{-0.2}^{+0.6}$  &$0.7_{-0.2}^{+0.3}$  &13.4 &(7), (6)\\ 
XX~Cha          &11:11:39.66 &$-$76:20:15.25&M1   &3705  &0.356&1.3 &-133.5&C&-9.07 & 0.49 &1.9 &$1.3_{-0.1}^{+1.1}$ &$1.4_{-0.4}^{+0.2}$ &28.1 &$2.4_{-0.1}^{+0.9}$  &$1.2_{-0.3}^{+0.2}$  &49.9 &(7), (6)\\ 
HM~Lup          &15:47:50.63 &$-$35:28:35.4 &M3   &3415  &0.181&0.0 &-115.3&C&-8.65 & 0.32 &2.0 &$1.1_{-0.1}^{+0.4}$ &$1.2_{-0.4}^{+0.3}$ &13.2 &$2.0_{-0.2}^{+0.8}$  &$0.2_{-0.1}^{+0.3}$  &10.6 &(8), (6)\\  
GW~Lup          &15:46:44.73 &$-$34:30:35.5 &M2   &3650  &0.292&0.1 &-90.3 &C&-8.32 &0.39  &1.6 &$4.2_{-0.5}^{+0.1}$ &$1.3_{-0.3}^{+0.2}$ &16.5 &$6.0_{-1.4}^{+0.0}$  &$0.7_{-0.2}^{+0.1}$  &45.3 &(8), (6)\\  
IM~Lup          &15:56:09.22 &$-$37:56:05.8 &M0   &3850  &1.993&0.2 &-5.7  &W&$<$-8.16 &0.43 &0.1 &$2.7_{-0.2}^{+0.2}$ &$1.2_{-0.2}^{+0.3}$ &18.5 &$4.2_{-0.5}^{+0.4}$  &$0.9_{ -0.1}^{+0.2}$  &75.4 &(8), (6), (9)\\  
Sz~73           &15:47:56.94 &$-$35:14:34.8 &M0   &3850  &0.405&3.6 &-97.2 &C&-7.98  &0.60 &2.5 &$2.7_{-0.2}^{+0.1}$ &$1.2_{-0.2}^{+0.2}$ &9.9  &$2.1_{-0.2}^{+0.7}$  &$0.9_{-0.4}^{+0.5}$  &13.4  &(8), (6)\\  
Sz~76           &15:49:30.74 &$-$35:49:51.4 &M1   &3705  &0.127&0.5 &-10.3 &W&$<$-10.17 &0.61 &14.3&$3.8_{-0.3}^{+0.4}$ &$1.4_{-0.5}^{+0.2}$ &8.6  &$3.0_{-1.8}^{+0.3}$  &$0.5_{-0.3}^{+0.5}$  &5.2  &(8), (6)\\  
Sz~96           &16 08 12.64 &$-$39 08 33.5 &M1.5 &3633  &0.548&0.3 &-11.0 &W&$<$-8.89  &0.39 &0.6 &$0.9_{-0.2}^{+0.2}$ &$0.9_{-0.2}^{+0.3}$ &15.4 &$2.5_{-0.1}^{+0.2}$  &$0.5_{-0.1}^{+0.2}$  &26.3 &(8), (6)\\

\hline
\end{tabular}
\tablebib{(1) \citet{2009A&A...497..379M}; (2) \citet{2008ApJ...687.1145S}; (3) \citet{2011ApJ...736..137S}; (4) \citet{2009ApJ...701.1188S}; (5) \citet{2008ApJ...680.1295S}; (6) \citet{2010A&A...520A..39O}; (7) \citet{1992ApJ...385..217G}; (8) \citet{1994AJ....108.1071H}; (9) \citet{2010A&A...517A..88W}}
\end{table*}
\normalsize

\end{appendix}
\bibliographystyle{aa}
\bibliography{references}

\end{document}